\documentclass[prd,aps,twocolumn,superscriptaddress,preprintnumbers,nofootinbib,showpacs]{revtex4}
\usepackage{graphicx}
\usepackage{exscale}
\usepackage[intlimits]{amsmath}
\usepackage{amsfonts}
\usepackage{amssymb,amscd}
\usepackage{epsfig}
\usepackage{xcolor}
\usepackage{pstricks}

\newcounter{commentdepth}
\newcommand{\be}{\begin{equation}}
\newcommand{\ee}{\end{equation}}

\renewcommand{\d}{\ensuremath\mathrm{d}}

\newcommand{\E}{\ensuremath\mathrm{e}}
\newcommand{\eps}{\ensuremath\varepsilon}

\newcommand{\norm}[1]{\ensuremath\left| #1 \right|}
\newcommand{\NN}{\ensuremath\mathbb{N}}

\newcommand{\LL}{\mathcal{L}}

\newcommand{\V}[1]{\boldsymbol{#1}}

\newcommand{\beqa}{\begin{eqnarray}}
\newcommand{\eeqa}{\end{eqnarray}}
\newcommand{\beq}{\begin{equation}}
\newcommand{\eeq}{\end{equation}}

\newcommand{\reftitle}[1]{}

\begin{document}
\preprint{\parbox[t]{\textwidth}
{\small May 26, 2008 \hspace*{5mm} \#\#\# \hfill ArXiv:0805.3804 [hep-ph]}}

\title{Nonperturbative Contributions to the QCD Pressure}

\author{Klaus~Lichtenegger}
\affiliation{Karl-Franzens-Universit\"at Graz, 8010 Graz, Austria}

\author{Daniel~Zwanziger}
\affiliation{New York University, New York, NY 10003, USA}

\begin{abstract}
\noindent We summarize the most important arguments why a
perturbative description of finite-temperature QCD is unlikely to be
possible and review various well-established approaches to
deal with this problem.
Then, using a recently proposed method,
we investigate nonperturbative contributions to the QCD pressure
and other observables (like energy, anomaly and bulk viscosity)
obtained by imposing a functional cutoff at the Gribov horizon.

Finally, we discuss how such contributions fit into the picture of
consecutive effective theories, as proposed by Braaten and Nieto,
and give an outline of the next steps necessary to improve this
type of calculation.

\end{abstract}


\maketitle
 
\section{Introduction}
\label{sec:solgap_intro}

\subsection{Rise and Fall of the Quark-Gluon Plasma}

\noindent One of the most striking properties of QCD is
\emph{asymptotic freedom}. For large momentum $p$, the coupling
$g(p)$ is small, so quarks and gluons can be treated
as if they were almost free particles --- in particular, they can be
treated with the sophisticated methods of perturbation theory.

While the situation is obviously different in the regime of low energies
(which is most relevant for nuclear physics), it was natural to expect
that a perturbative decription could be applied to QCD at sufficiently
high temperatures. After all, high temperature $T$ implies high average
particle momentum and thus a small coupling, i.e. almost free particles.

For this scenario, the term ``quark-gluon plasma'' was
coined~\cite{Shuryak:1978ij}, and one could expect a phase
transition where, on a certain curve in the $\mu$-$T$-space (where
$\mu$ denotes the chemical potential), hadrons melt into
such a plasma.

This phase transition offered a natural solution to a problem posed by
Hagedorn, \cite{Hagedorn:1965st}, who found that due to an exponential
increase of the number of accessible states, the temperature of a hadron
could not exceed a certain limit $T_{\mathrm{H}}\approx160\text{MeV}$.

The picture of hadrons melting into a plasma of (almost) free quarks and
gluons, however, turned out to be too naive. In principle, this should have
been clear at least since 1980, when it was
shown~\cite{Linde:1980ts, Gross:1980br} that at order $g^6$ a natural
barrier arises for any perturbative description. Even earlier than that,
the simple fact that the infinite-temperature limit of four-dimensional
Yang-Mills theory is a three-dimensional \emph{confining} Yang-Mills
theory could and should have been regarded as a sign that any straightforward
perturbative approach to high-temperature QCD was necessarily doomed.

It took however more than 20 years until it (slowly) began to be accepted
that the high-temperature phase of QCD has little to do with a
conventional plasma. The results of the RHIC experiments,~\cite{Shuryak:2004cy},
showed clearly that also above the phase transition, bound state phenomena
can not be neglected, and the description as a perfect fluid is much more
accurate than the one as a weakly interacting plasma.

While the term ``quark-gluon plasma'' is still widely used, one begins
to speak (more accurately, though somehow using an oxymoron) of a
``strongly coupled quark gluon plasma''~\cite{Bannur:1998nq, Shuryak:2006vg},
or even a ``quark-gluon soup''.
 
With the experimental results which are --- for certain observables --- an
order of magnitude away from the predictions 
for a weakly coupled plasma (see for example data on the elliptic flow
in~\cite{Muller:2006ee}) an accurate description of the high-temperature phase
remains a challenge for theoretical physics. One conclusion, however,
seems to be clear: In the high-$T$ regime, perturbation theory has to be
replaced or at least supplemented by nonperturbative methods.

\subsection{Organization of paper}
\label{ssec:solgap_organization}

\noindent After the introduction given in~\ref{sec:solgap_intro},
in~\ref{sec:solgap_highTQCD} we briefly review aspects
of finite-temperature QCD. In particular, in~\ref{ssec:solgap_howstudy}
we discuss which thermodynamic quantities might be interesting
to look at, while in~\ref{ssec:solgap_threescales}
we examine the arguments for a breakdown of perturbation theory.
In~\ref{ssec:solgap_pertappr} we summarize the known
perturbative results, which can be rederived and extended by
effective field theory methods, which are discussed
in~\ref{ssec:solgap_effField}. In~\ref{ssec:solgap_functional}
we discuss previous functional approaches and
in~\ref{ssec:solgap_lattice} lattice results.
In~\ref{ssec:solgap_compare} we compare the results of these
different methods and discuss questions of convergence.

In~\ref{sec:solgap_semipert} we approach the problem
with a new ``semi-perturbative" method \cite{Zwanziger:2006sc, Zwanziger:2007zz, Zwanziger:2006wz} which is
briefly reviewed in~\ref{ssec:solgap_eqstatelocal}.  The physics behind this method is that the functional cut-off at the Gribov horizon suppresses the infrared components of the gluon field \cite{Gribov:1977wm}, so that the infrared divergences of finite-temperature field theory found by Linde \cite{Linde:1980ts} do not arise \cite{zahed:1999}.
This method typically involves a temperature-dependent
renormalization scale, an issue we discuss
in~\ref{ssec:solgap_renormscale}. In~\ref{ssec:solgap_calcmeth}
we examine the calculational methods used to solve the resulting
equations and the expansion used to extract the asymptotic form,
before we present our results in~\ref{ssec:solgap_results}.

In section~\ref{sec:solgap_discussion} we discuss these results
and how they are related to other approaches.
In particular, in~\ref{ssec:solgap_access_nonpert} we compare
different ways to access the nonperturbative sector of hot QCD,
in~\ref{ssec:solgap_convseries} we resume the discussion
of convergence and in~\ref{ssec:solgap_furthersteps} we
present some ideas about how to pursue further research.

In~\ref{klsec:solgap_summary} we summarize our results
and give a brief outlook.

\section{High Temperature QCD}
\label{sec:solgap_highTQCD}

\subsection{How to Study High Temperatures}
\label{ssec:solgap_howstudy}

\noindent It is useful to rescale thermodynamic quantities with appropriate
powers of the temperature. In particular, the free energy per unit volume,\footnote{In statistical mechanical usage the ``free energy" is given by $F = - wVT$.} the pressure, and the energy per unit volume
\beq
  w = \frac{\ln Z}{V}\,,\qquad p = \frac{w}{\beta}\,,\qquad e = - \frac{\partial w}{\partial\beta}
  \label{eq:solgap_thermoform}
\eeq
are rescaled to
\beq
  w_{\text{r}} = \frac{w}{T^3}\,,\qquad p_{\text{r}} = \frac{p}{T^4}\,,\qquad e_{\text{r}} = \frac{e}{T^4}\,.
\eeq

\noindent The \emph{anomaly} $A=e-3p$ is rescaled to
\beq
  A_{\text{r}} = \frac{A}{T^4} = \frac{e-3p}{T^4}\,.
  \label{eq:solgap_Ar_def}
\eeq

According to~\cite{Kharzeev:2007wb}, up to a perturbative contribution,
the \emph{bulk viscosity} $\zeta$ for hot gauge theories is given by the
logarithmic derivative of the anomaly,
\beq
  \zeta = \frac1{9\omega_0} \left\{ T^5\,\frac{\partial}{\partial T} \left( \frac{e-3p}{T^4} \right)
  + 16\,\left|\eps_{\text{V}}\right| \right\}\,,
  \label{eq:solgap_viscos_kharzeev}
\eeq
where $\omega_0$ denotes a perturbative scale and $\eps_{\text{V}}$
is a perturbative contribution. This formula can be derived from the Kubo
formula of linear response theory. That the viscosity is linear in the trace
of the energy-momentum tensor $\Theta_{\mu\nu}$ (instead of quadratic)
is not surprising in view of the Schwinger-Dirac relations, as discussed for
example in~\cite{BrownDiracSchwinger}.

\subsection{The Perturbative Problem in the Infrared}
\label{ssec:solgap_threescales}

\noindent Perturbative calculations at finite temperature are dramatically
different from those at $T=0$. One of the most striking differences
is that one cannot determine the order of a graph by simply
counting the number of vertices. Actually a vacuum or propagator
graph may be nonanalytic in $g^2$.

While ultraviolet divergences are regulated exactly the same way
as in the zero-temperature theory, with no additional effort necessary,
for $T>0$ additional \emph{infrared} divergences appear.  They come from Matsubara frequency $n = 0$, which has the infrared divergences of 3-dimensional Euclidean gauge theory that are even more severe than in 4 dimensions.  For this reason the Gribov horizon, which affects primarily infrared components of the gauge field, is more important at finite and high $T$ than at $T = 0$.  

Here, however, another subtlety of thermal field theory comes to rescue:
Thermal fluctuations give rise to self-energy, which, in the static limit
$\V{p}\to\V{0}$ corresponds to a mass $m$. At first glance, there are two
natural candidates for the scale of such a mass, the electric screening mass
$m_{\mathrm{el}} \sim gT$ and the magnetic screening mass
$m_{\mathrm{mag}} \sim g^2T$.

\begin{figure}
\begin{center}
  \includegraphics[width=8cm]{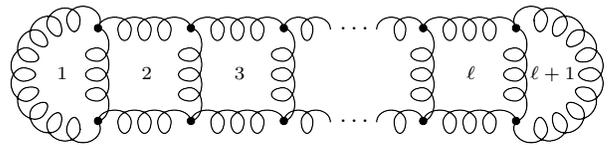}
  \put(-100,8){$\hdots$} \put(-100,43){$\hdots$}
  \put(-207,24){\scriptsize $1$}
  \put(-175,24){\scriptsize $2$}
  \put(-140,24){\scriptsize $3$}
  \put(-52,24){\scriptsize $\ell$}
  \put(-28,24){\scriptsize $\ell+1$}
\end{center}
\caption{A ``ladder diagram'' contributing at order $g^6$ to the free
  energy for $\ell\ge 3$.}
\label{fig:solgap_ladder}
\end{figure}

The mass which is dynamically generated appears in the value for ladder diagrams
like the one depicted in fig.~\ref{fig:solgap_ladder} (see section 8.7
of~\cite{Kapusta:2006pm}). For this type of diagram we obtain (ignoring
the complicated tensorial structure)
\beq
  I_{\ell} \sim \begin{cases} g^{2\ell}T^4 & \text{for}\;\ell = 1,\,2 \\[6pt]
  g^6\,T^4\,\ln \frac{T}{m} & \text{for}\;\ell = 3 \\[6pt]
  g^6\,T^4\,\left( \frac{g^2T}{m}\right)^{\ell-3} & \text{for}\;\ell > 3\,.
  \end{cases}
\eeq

If $m$ were independent of $g$ or, like $m_{\mathrm{el}}$, of order $gT$, we could
proceed with perturbation theory without serious problems, since an increasing
number of loops would always correspond to an increasing power of the coupling $g$.
It turns out, however, that $m$ is (in the best case) of the order of the
\emph{magnetic} screening mass, $m_{\mathrm{mag}} \sim g^2 T$.

Thus for any value $\ell\ge 3$ one has contributions of order $g^6$,
the perturbative procedure becomes impracticable unless a suitable
resummation technique is available --- and such a technique has not been
found up to now.

\subsection{Direct Perturbative Approach}
\label{ssec:solgap_pertappr}

\noindent We have seen that the perturbative treatment of the QCD free energy 
runs into fundamental problems at order $g^6$. Still, one can expect that for
sufficiently small values of $g$ (i.e. for sufficiently high temperatures) the
possible perturbative description (to order $g^5$) still provides a good
description.

This is indeed the case (although, as we will see in section~\ref{ssec:solgap_compare},
only at ridiculously high temperatures). Unfortunatey, even these calculations turn
out to be highly involved. We summarize here known results, which
are also collected in~\cite{Kapusta:2006pm}, but specialize them to the
case of pure gauge theory.

Zeroth order just gives the Stefan-Boltzmann law for SU(N) gauge theory,
\begin{align}
  \frac{p^{(0)}}{T^4} &= - (N^2-1)\,\frac{\pi^2}{45}\,.
\end{align}
For the second-order contribution, one obtains,~\cite{Shuryak:1977ut, Chin:1978gj},
\begin{align}
  \frac{p^{(2)}}{T^4} &= - (N^2-1)\,\frac{\pi^2}9\,C_A\,\left( \frac{g}{4\pi} \right)^2
\end{align}
with $C_A$ denoting the Casimir of the adjoint representation, and $C_A = N$ for SU(N). Due to non-analyticity,
one has a contribution of $\mathcal{O}(g^3)$, calculated in~\cite{Kapusta:1979fh},
\begin{align}
  \frac{p^{(3)}}{T^4} &=   (N^2-1)\,\frac{\pi^2}9\,C_A^{3/2}\,\frac{16}{\sqrt3}\,\left( \frac{g}{4\pi} \right)^3\,.
\end{align}
The $g^4\,\ln g$ contribution has been calculated in~\cite{Toimela:1982hv},
the full $g^4$ term has been obtained in~\cite{Arnold:1994ps, Arnold:1994eb}
\begin{align}
  \frac{p^{(4)}}{T^4} &= (N^2-1)\,\frac{\pi^2}9\,C_A^2\,\left( \frac{g}{4\pi} \right)^4\,
    \Bigg\{ 24\,\ln\left( \frac{C_A}{3}\,\frac{g}{2\pi}\, \right) \nonumber \\
    &\qquad - \Big[ \frac{22}3\,\ln\frac{\mu(T)}{2\pi T} +\frac{38}3\,\frac{\zeta'(-3)}{\zeta(-3)} \nonumber \\
    &\qquad\qquad  -\frac{148}3\,\frac{\zeta'(-1)}{\zeta(-1)} - 4\,\gamma_E + \frac{64}5 \Big] \Bigg\} 
\end{align}
where $\gamma_{\mathrm{E}}$ denotes the Euler-Mascheroni constant and $\zeta$
the Riemann zeta function.

At order $g^5$, one obtains,~\cite{Zhai:1995ac}
\begin{align}
    \frac{p^{(5)}}{T^4} &= (N^2-1)\,\frac{\pi^2}9\,
       \left( \frac{g}{4\pi} \right)^5\,\sqrt{\frac{C_A}3}\,C_A^2\, \nonumber \\
    &\quad  \cdot \Big[ 176\,\ln\frac{\mu(T)}{2\pi\,T}+176\,\gamma_E \nonumber \\
    &\qquad - 24\,\pi^2 + 494 + 264\,\ln 2 \Big].
\end{align}

\subsection{Effective Field Theory}
\label{ssec:solgap_effField}

\noindent The result of order $g^5$ is the last one
obtained in strict perturbation theory. It has been rederived by Braaten
and Nieto~\cite{Braaten:1995cm}, using an effective field theory method
that is built on the idea of dimensional reduction~\cite{Appelquist:1981vg,
Braaten:1994na}.

The problem of infrared divergences is adressed by two effective theories
that are constructed ``below'' perturbative QCD. We know that there are
three important scales present, namely

\begin{center}
\begin{tabular}{rcl}
  $2\pi T$ & \ldots & scale of ``hard modes'' \\
  $g\,T$ & \ldots & chromoelectric scale \\
  $g^2\,T$ & \ldots & chromomagnetic scale.
\end{tabular}
\end{center}

\noindent Thus it makes sense to describe each scale in a somewhat
different way. To do this, two cutoff scales $\Lambda_E$ and $\Lambda_M$
are introduced, that have to satisfy
\beq
  2\pi T \gg \Lambda_E \ge gT \gg \Lambda_M \ge g^2T\,.
\eeq
The region with $p>\Lambda_E$ can be reliably described by perturbative
QCD, and for this contribution to the free energy, called $f_E$, one obtains a power
series in $g^2$ with coefficients that can depend on $\ln\frac{T}{\Lambda_E}$.

For $\Lambda_E>p>\Lambda_M$, with the hard modes integrated
out, an effective three-dimensional theory, called \emph{electrostatic QCD}
(EQCD) is introduced,
\begin{align}
  \LL_{\text{EQCD}} &= \frac14 F^a_{ij}F^a_{ij} + \frac12 (D_iA_0)^a(D_iA_0)^a \nonumber \\
    &\quad+ \frac12m_E^2A_0^aA_0^a + \frac18\lambda_E\,(A_0^aA_0^a)^2 \nonumber \\
    &\quad+ \delta\LL_{\text{EQCD}}\,,
\end{align}
where $F^a_{ij}=\partial_i A^a_j -\partial_j A^a_i+g_E\,f^{abc}A^b_iA^c_j$
denotes the magnetostatic field strength tensor and $\delta\LL_{\text{EQCD}}$
contains all other local (3-dimensionally) gauge-invariant operators of dimension
three or higher that can be constructed from $A_i$ and $A_0$.
The parameters $g_E$, $m_E$, $\lambda_E$ are detemined by matching to
perturbative QCD, in particular one has $m_E \sim m_{\text{el}} \sim gT$.

This theory still allows perturbative treatment, making use of an expansion in the
dimensionless quantities $\frac{g_E^2}{m_E}\sim g$, $\frac{\lambda_E}{m_E}$ etc.
This gives for the contribution $f_M$ to the free energy a power series in $g$,
with coefficients that depend on $\ln\frac{\Lambda_E}{gT}$ and $\ln\frac{gT}{\Lambda_M}$.
The whole series is multiplied by the common factor $(gT)^3\,T$.

The infrared cutoff $\Lambda_M$ of EQCD is the UV cutoff of another
theory, \emph{magnetostatic QCD} (MQCD),
\begin{align}
  \LL_{\text{MQCD}} &= \frac14 F^a_{ij}F^a_{ij} + \delta\LL_{\text{MQCD}}\,,
\end{align}
with $\delta\LL_{\text{MQCD}}$ denoting all gauge-invariant operators of dimension $5$
or higher. This theory is confining and thus truly nonpertubative, but according
to~\cite{Braaten:1995cm}, this contribution to the free energy, called $f_G$, can still be
expanded in a power series in $g$, which is multiplied by a general factor $(g^2T)^3$.
However the value of the coefficient cannot be determined perturbatively.

Since the (well-established) nomenclature may seem slightly misleading at
first glance, we have tried to give a graphical representation in
fig.~\ref{fig:solgap_sketchscales}.

\begin{figure}
\begin{center}
\begin{picture}(200,90)
  \put(20,30){\color{gray}\line(1,0){180}}
  \put(20,60){\color{gray}\line(1,0){180}}
  \put(0,28){$\Lambda_M$}
  \put(0,58){$\Lambda_E$}
  \put(25,13){$g^2T$}
  \put(25,43){$gT$}
  \put(25,73){$2\pi T$}
  \put(57,13){\color{gray}MQCD}
  \put(57,43){\color{gray}EQCD}
  \put(57,73){\color{gray}pQCD}
  \put(100,13){$\displaystyle f_G=(g^2T)^3\sum_{k=0}^{\infty} c_k\,g^k$}
  \put(100,43){$\displaystyle f_M=(gT)^3\sum_{k=0}^{\infty} b_k\,g^k$}
  \put(100,73){$\displaystyle f_E=T^3\sum_{k=0}^{\infty} a_k\,(g^2)^k$}
\end{picture}
\end{center}
\caption{The scales of perturbative QCD (pQCD), electrostatic QCD (EQCD),
magnetostatic QCD (MQCD) and the different contributions $f_E$, $f_M$
and $f_G$ to the free energy. The coefficients $a_k$ and $b_k$ are polynomials
in logarithms of ratios of scales, $a_k=P_k(\ln\frac{T}{\Lambda_E})$,
$b_k=Q_k(\ln\frac{\Lambda_E}{gT},\,\ln\frac{gT}{\Lambda_M})$. While the
coefficients $a_k$ and $b_k$ can be determined, at least in principle,
in perturbation theory, this is not possible for $c_k$.}
\label{fig:solgap_sketchscales}
\end{figure}

\medskip

MQCD is genuinely nonperturbative, its degrees of freedom are $(2+1)$-dimensional
glueballs. In~ \cite{Braaten:1994na} it was suggested to calculate the contributions
from this scale directly by lattice methods.

With the effective field theory, it is possible to compute the
$g^6\,\ln g$ contribution~\cite{Kajantie:2002wa}. The contribution
obtained this way has to be regarded as partly conjectural, since
the argument inside the logarithm is not clearly defined until the full
$g^6$ contribution is known.

The result thus relies on a supposed structure of cancellation patterns. In addition, it
is believed to be reliable only for sufficiently high temperatures (which could, however,
mean down to $T\approx 2T_C$), since description by a three-dimensional theory is valid
only for such temperatures . With these caveats in mind, one obtains for pure
$SU(3)$ gauge theory
\begin{align}
     \frac{p^{(6)}}{T^4} &= \frac{8\pi^2}{45}\,\left( \frac{g^2}{4\pi^2}\right)^3
     \Bigg\{ \left[ -659.2 + 742.5\,\ln\frac{\mu(T)}{2\pi T} \right]\,\ln\frac{g^2}{4\pi^2} \nonumber \\
     &\qquad - 475.6\,\ln\frac{g^2}{4\pi^2} - \frac{1815}{16}\,\ln^2\frac{\mu(T)}{2\pi T} \nonumber \\
     &\qquad+ 2932.9\,\ln\frac{\mu(T)}{2\pi T} + q_c^{(0)} \Bigg\}
\end{align}
with a yet undetermined coefficient $q_c^{(0)}$ for the pure $g^6$ contribution.
(See also~\cite{Schroder:2002re, Schroder:2003uw}.)

This coefficient consists of both perturbative contributions (from pQCD
and EQCD) and nonperturbative contributions (from MQCD). It was
estimated in~\cite{Kajantie:2002wa} by a fit to four-dimensional lattice data
for the pressure.\footnote{The problem with such a procedure is that in the regime
where lattice data is available, the contributions of higher order may also be large.}

Some of the perturbative contributions of order $g^6$ are known by
now~\cite{Laine:2005ai, Schroder:2004gt},
but others remain unknown. The nonperturbative coefficient has been determined
by three-dimensional lattice calculations and matching to perturbative four-loop
calculations in~\cite{Hietanen:2004ew}
(see also~\cite{Hietanen:2006rc} for some cases with $N\ne 3$)
and~\cite{DiRenzo:2006nh}. One obtains
\beq
  w_{\text{np}}^{(6)} = g_3^6\,\frac{(N^2-1)\,N^2}{(4\pi)^4}\,B_G
  \label{eq:solgap_lattice_npcoeff1}
\eeq
with $g_3^2=g^2T(1+\mathcal{O}(g^2))$ and the constant
\beq
   B_G = -0.2 \pm 0.4^{\text{MC}} \pm 0.4^{\text{SQ}}
  \label{eq:solgap_lattice_npcoeff2},
\eeq
where the first error stems from the Monte Carlo simulation, the second
one from the Stochastic Quantization procedure employed to obtain the final
result. Note that $B_G=0$ is compatible with this result.

\subsection{Functional Approaches}
\label{ssec:solgap_functional}

\noindent Due to the limitations of perturbation theory, nonperturbative methods
definitely deserve a closer look --- moreover the effective field theory approach
also relies on the ability to calculate certain quantities nonperturbatively.

As in the zero-temperature case~\cite{Alkofer:2000wg},
also for finite temperature, fundamental aspects of Yang-Mills theory and
QCD are accessible to functional methods based on Dyson-Schwinger
equations (DSEs),~\cite{Maas:2004se, Maas:2005hs}.

For certain asymptotic situations (deep ultraviolet, deep infrared,
infinite temperature limit) several analytic results can be obtained;
but in general numerical studies of truncated DSE systems are necessary.

In addition to the standard truncations, finite temperature calculations
require also some treatment of the Matsubara series. Usually it is replaced
by a finite sum, even though this means that the limit of four-dimensional
zero-temperature theory is now technically hard to access.

At the present level, these restrictions make it difficult to obtain
precise quantitative results. Nevertheless there is reasonable confidence
about the qualitative picture that arises from these studies. Both from
infrared exponents and from numerical results one sees that the
soft modes are not significantly affected by the presence of hard
modes, thus the confining property of the theory cannot be
expected to be lost in the high-temperature phase.

Consequently, while the over-screening (which would attribute an infinite
amount of energy to free color charges) of chromoelectric gluons is reduced to
screening (as it is the case for electric charges in a conventional plasma),
chromomagnetic gluons remain over-screened and thus confined, which
renders any description of such gluons as almost free (quasi-)particles
meaningless.

While the functional method yields considerable insight into propagators
and related quantities, unfortunately the pressure (and quantities derived
from it) are, up to now, difficult to access in this approach. Nevertheless the
results obtained so far by functional methods provide additional evidence
for the picture of bound states playing an important role even at very high
temperature and part of the gluon spectrum (the chromomagnetic sector)
being confined at any temperature.

\subsection{Lattice Gauge Theory}
\label{ssec:solgap_lattice}

\noindent Lattice Gauge Theory is generally considered the most rigorous approach
to nonperturbative QCD, and so it is natural to also study thermodynamics
on the lattice.

The drawbacks of the method, however, are known as well: To reliably
approach the thermodynamic  and the continuum limits, extrapolations which
require calculations with various different lattice sizes are necessary.
The inclusion of fermions is expensive, especially if good chiral properties
are required.

\begin{figure}
\begin{center}
  \includegraphics[width=8cm]{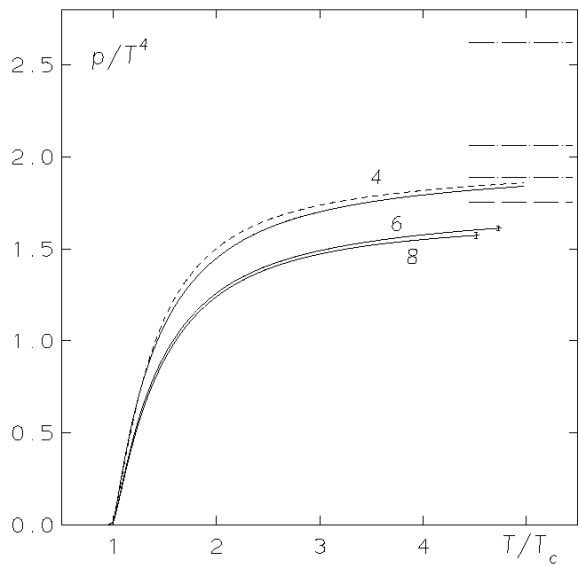}
\end{center}
\caption{Rescaled pressure of SU(3) lattice gauge theory,
  from~\cite{Boyd:1996bx}, where  $T_c$~is the transition
  temperature, and $N_\tau = 4, 6, 8$.}
\label{fig:solgap_latt_pressure}
\end{figure}

\begin{figure}
\begin{center}
  \includegraphics[width=8cm]{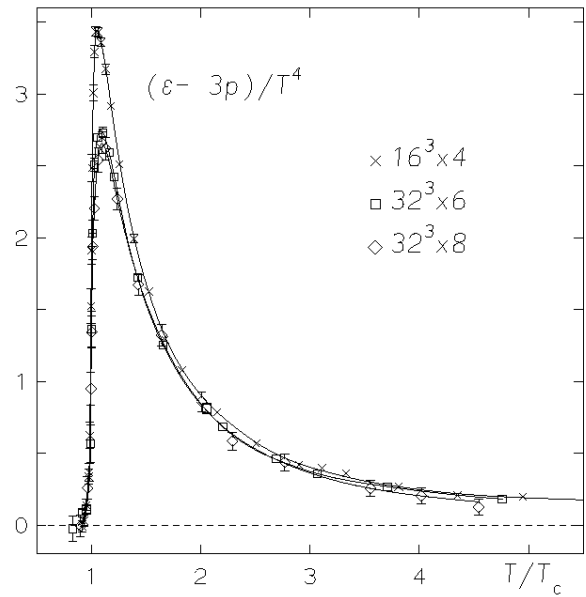}
\end{center}
\caption{Rescaled anomaly of SU(3) lattice gauge theory,
  from~\cite{Boyd:1996bx}}
\label{fig:solgap_latt_anomaly}
\end{figure}

Despite these drawbacks, lattice data is (apart from possible experiments) the thing
one usually compares any other calculation to. For pure SU(3) gauge theory, the problem
of determining the equation of state is regarded as solved since the publication
of~\cite{Boyd:1996bx}, where results are further confirmed
in~\cite{Okamoto:1999hi}. (Note, however, that there remain certain doubts about
the accuracy of the infinite-volume limit, see~\cite{Gliozzi:2007jh}.) The results for the
pressure and the anomaly are displayed in figs.~\ref{fig:solgap_latt_pressure}
and~\ref{fig:solgap_latt_anomaly}.

\subsection{Other approaches}

\noindent Pisarski~\cite{Pisarski:2006hz} has observed from lattice data~\cite{Boyd:1996bx}
that $(e - 3p)/T^4 \times T^2$ is approximately constant in a broad range above
$T_c$. From this and physical reasoning he obtained the formula
\beq
  p_{\rm pure\;glue}(T) \approx f_{\rm pert}(T^4 - T_c^2 T^2).
\eeq
It is notable that no $T^3$ correction is apparent.

An active area of research is the AdS/CFT duality~\cite{Maldacena:1997re} and
AdS/QCD duality~\cite{Aharony:1999ti}, including in particular duality at finite
temperature (for a pedagogical introduction see~\cite{Peeters:2007ab}).
In this connection it is interesting to note that a formula similar to Pisarski's has
recently been obtained from this duality and the (truncated) entropy density of
the horizon of a deformed Euclidean $\text{AdS}_5$ black
hole,~\cite{Andreev:2007zv}.

\subsection{Comparison of Results}
\label{ssec:solgap_compare}

\noindent Knowing that perturbation theory is limited to some fixed order in $g$,
we can still estimate how good the possible perturbative description
actually is. Ways to judge this are to check whether contributions from
higher orders are small compared to those from lower orders or to compare
perturbative expressions to results of lattice calculations.

\begin{figure}
  \includegraphics[width=8cm]{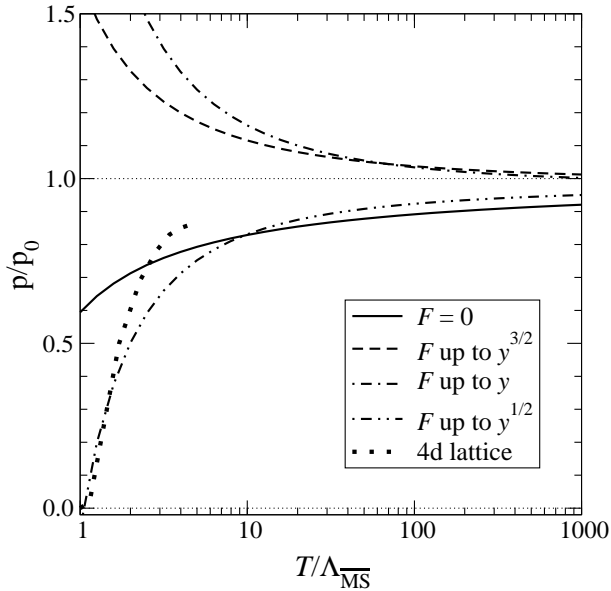}
  \caption{The convergence of an (optimized) perturbation series for
  ``long-distance contribution'' to the pressure, from~\cite{Kajantie:2000iz}.
  The order of the expansion is characterized by the dimensionless
  parameter $y\sim\frac{g^2T^2}{g_3^4}$, where $g_3^2$ denotes the
  gauge coupling in the effective three-dimensional theory. The
  perturbative contribution $F$ to the free energy contains a factor
  $\left(\frac{g_3^2}T\right)^3$, inclusion of lower powers of $y$
  corresponds to higher orders in perturbation theory.}
  \label{fig:solgap_pertkajantie}
\end{figure}

Unfortunately, both methods suggest that the convergence is extremely
poor for temperatures of the order of several $T_c$, where $T_c$ is the
transition temperature, and to obtain good convergence one has to look
at least at the electroweak scale,~\cite{Zhai:1995ac, Kajantie:2002wa}.
A plot of the results of optimized perturbation theory are given in
fig.~\ref{fig:solgap_pertkajantie}.

It has been conjectured,~\cite{Laine:2003ay}, that the results of order
$g^6$ are not significantly changed by higher orders (since one can hope
to have obtained at order $g^6$ the main contribution from each scale;
perhaps also due to the fact that originally large terms of higher orders
cancel against each other). To the knowledge of the authors, however,
there is no strong evidence to support this conjecture.

From the existing data one cannot even exclude the unsettling possibiliy
that for ``physical'' temperatures the perturbation series already begins to
diverge at some order $n \le 6$. This would mean that contributions from
higher orders are of comparable magnitude to those of low order
and no systematic cancellations occur. If this were indeed the case, we could
not expect to have any reliable perturbative description for temperatures
which are accessible in current experiments.

\medskip

\noindent One should mention that there is an additional ambiguity in the perturbative
results. All terms beyond the Stefan-Boltzmann contribution contain some power of
the running coupling $g$. Thus, for all practical calculations there is some dependence
on the scale $\mu$, at which $g(\mu)$ is evaluated. Traditionally, one chooses
$\mu(T)=2\pi T$ in the high-temperature regime, but there are alternative approaches,
for example application of the \emph{principle of minimal sensitivity},~\cite{Stevenson:1981vj,
Blaizot:2003iq, Inui:2005gu}. We will discuss this question in more detail in
section~\ref{ssec:solgap_renormscale}.

\section{The Semiperturbative Approach}
\label{sec:solgap_semipert}

\subsection{Equation of State from a Local Action}
\label{ssec:solgap_eqstatelocal}

\noindent An alternative approach that combines nonperturbative
elements with perturbative expansions has been developed
in~\cite{Zwanziger:2006sc, Zwanziger:2007zz, Zwanziger:2006wz},
which we now briefly describe.

The basic physical idea is that the infrared divergences of finite-temperature perturbation theory do not arise when the domain of functional integration is cut-off at the Gribov horizon.  The cut-off will be done in Coulomb gauge which is well adapted to finite-temperature calculations.  Indeed both the gauge condition, $\partial_i A_i({\bf x}, t) = 0$, and the cut-off at the Gribov horizon are applied to 3-dimensional configurations on each time slice~$t$, and are entirely independent of the temporal extent of the lattice $ 0 \leq t \leq \beta $, where $\beta = 1/kT$.  

The functional cut-off at the Gribov horizon is effected  at first by adding a non-local term $S_{NL}(A)$ to the action \cite{Zwanziger:1989mf, Zwanziger:1993}.  The non-local term then gets replaced by a local, renormalizable term $S_L$ in the action by means of an integration over a multiplet of auxiliary fermi and bose ghost pairs ,
\beq
\exp[- S_{NL}(A)] = \int \d\varphi \d\bar{\varphi} \d\omega \d\bar\omega \ \exp[- S_L(\varphi, \bar\varphi, \omega, \bar\omega)].
\eeq
The BRST symmetry is explicitly broken by this term, an effect which, alternatively,
may be interpreted as spontaneous BRST breaking~\cite{Maggiore:1993wq}.  Although the breaking of BRST invariance precludes the definition
of observables as elements of the cohomology of the BRST-operator, the eqivalence to the canonical formulation has been established~\cite{Zwanziger:2006sc}, thereby ensuring the physical foundation of the approach, including unitarity.  Here the physicality of the Coulomb gauge plays an essential role.

The new term in the action depends on a mass parameter $m$ which appears in the Lagrangian density 
\beqa
\label{Lgamma}
{\cal L}_m & = &  - {m^4 \over 2N g^2} 
(D-1)(N^2 -1)
\\    \nonumber
& & +{m^2 \over (2N)^{1/2} g} \ 
[ \  D_i(\varphi -  \bar\varphi)_i
+  g(D_i c \times \bar\omega_i) \ ]^{aa}.   
\eeqa 
The adjoint part of the Bose ghost $(\varphi - \bar\varphi)_i$  mixes with the gauge field $A_i$ through the term
$D_i(\varphi -  \bar\varphi)_i = (\partial_i + g A_i \times)(\varphi -  \bar\varphi)_i$.  At tree level one obtains a gluon propagator, 
\beq
D = { 1 \over k_0^2 + E^2(\V{k}) },
\eeq
that satisfies the Gribov dispersion relation
\beq
  E(\V{k}) = \sqrt{\V{k}^2 + \frac{m^4}{\V{k}^2}}
  \label{eq:solgap_gribovdisp}.
\eeq

The functional cut-off at the Gribov horizon imposes the condition that the free energy $W$ or quantum effective action $\Gamma$ be stationary with respect to $m$,
\beq
\label{horizcond}
{ \partial W \over \partial m } = 
- { \partial \Gamma \over \partial m } = 0.
\eeq	
This ``horizon condition" has the form of a non-perturbative gap equation that determines the Gribov mass~$m = m(T, \Lambda_{QCD})$, and thereby provides a new vacuum, around which a perturbative expansion is again possible. 

	The most powerful non-perturbative methods available are called for to solve this system.  However in the present work we shall modestly investigate a semi-perturbative method~\cite{Zwanziger:2006sc}, in which one calculates all quantities perturbatively in $g$, including $\Gamma$, taking $m$ to be a quantity of order $g^0$, and then one substitutes for $m$ the non-perturbative solution to the gap equation (\ref{horizcond}).  We shall find that this method can be a good approximation only at extremely high energies.  Nevertheless as a matter of principle, it is a significant success that for thermodynamic observables this procedure gives finite results
precisely at the order, $g^6$ at which ordinary perturbation theory diverges.

\subsection{The Gap Equation}
\label{ssec:solgap_gapeq}

\noindent In lowest non-trivial order in the semi-perturbative
method~\cite{Zwanziger:2006sc}, the gap equation (\ref{horizcond}) reads
after separation in an $m^*$-dependent and a $T$-dependent part
\begin{align}
  f \left( m^* \right) &= y(T)
    \label{eq:solgap_gapeq} \\
  f \left( m^* \right) &:= \frac12\ln\frac1{m^*} + \int_0^{\infty} \frac{\d x}{u(x)}\,\frac1{\E^{m^*u(x)}-1} 
    \label{eq:solgap_gapeqLHS} \\
  y(T) &:= \frac{3\pi^2}{N\,g^2(\mu)} - \frac14\,\ln\frac{\E\mu^2(T)}{2\,T^2}\,.
     \label{eq:solgap_gapeqRHS} 
\end{align}
where $m^* \equiv m_r = m/T$ is the rescaled Gribov mass and
\beq
  u(x) \equiv \sqrt{x^2+\frac1{x^2}}\,
  \label{eq:solgap_III_reddisprel}
\eeq
is the reduced dispersion relation.  An important source of ambiguity,
shared with other (semi)perturbative approaches, is the choice of
the temperature-dependent scale $\mu(T)$ at which the coupling
$g$ is evaluated.

\subsection{Choice of the Renormalization Scale}
\label{ssec:solgap_renormscale}

\noindent We consider the coupling $g^2(\mu)$ at some renormalization
scale $\mu(T)$. For a certain temperature $T$, the optimal renormalization
scale should be chosen equal to the scale that governs the behaviour of the
system. For field theory at high temperatures, this scale is expected
to be equal to the lowest Matsubara frequency, i.e. $2\pi T$; for small
$T$ it should be constant.

Since we are considering a confining theory with a
mass gap, for low temperatures the optimal renormalization scale is not expected
to go to to zero. For a system at very low (even zero) temperature, the most
characteristic scale is not the very small average kinetic energy, but instead the mass
of the lightest physical object, which is some bound state (a hadron in full
QCD, a glueball in pure gauge theory). Actually, as long as we are in the
confining region (i.e. below $T=T_{\mathrm{c}}$), the mass of bound states
will always be ``more important'' than the thermal energy.

These restrictions, together with some conditions of ``naturalness'',
can be summarized by demanding that  the renormalization scale
$\mu(T)$ should fulfill:
\[
 \begin{array}{clcl}
    \text{(I)} &\mu(T) \approx \mu_0=\text{const} & & \text{for}\;T \ll T_{\mathrm{c}}, \\[3pt]
    \text{(II)} & \mu(T) \approx 2\pi T && \text{for}\;T\gg T_{\mathrm{c}}, \\[3pt]
    \text{(III)} & \text{continuous}, && \\[3pt]
    \text{(IV)} & \text{monotonically rising} && \text{for all}\;T, \\[3pt]
    \text{(V)} &  \text{convex} && \text{for all}\;T
  \end{array}
\]
Conditions (I) and (II) might be replaced or supplemented by the
asymptotic conditions
\beq
 \begin{array}{cl}
    \text{(I')} & \displaystyle \lim_{T\to 0} (\mu(T)-\mu_0) = 0\,, \qquad
    \lim_{T\to 0} \frac{\mu(T)-\mu_0}{T} = 0\,, \\[9pt]
    \text{(II')} & \displaystyle \lim_{T\to \infty} T^n \left( \mu(T)-2\pi T \right) = 0
      \quad \text{for all}\;n\in\NN_0
  \end{array}
\eeq
Due to the phase transition at
$T=T_\mathrm{c}\approx\Lambda_{\mathrm{QCD}}$,
a simple choice is
\beq
   \mu(T) = \begin{cases} \mu_{0,\mathrm{l}}(T) := 2\pi\Lambda_{\overline{MS}} & \mathrm{for}\; T<\Lambda_{\overline{MS}} \\
    \mu_{0,\mathrm{h}}(T) := 2\pi T & \mathrm{for}\; T\ge \Lambda_{\overline{MS}}\,.
   \end{cases}
   \label{eq:solgap_muTdirect}
\eeq
This choice is supported by the fact that $2\pi\Lambda_{\overline{MS}}$
is in the order of magnitude of glueball masses. Another reasonable choice is
\beq
  \mu(T) = 2\pi T + 2\pi\,\Lambda_{\overline{MS}}\,\E^{-T/\Lambda_{\overline{MS}}}\,.
   \label{eq:solgap_muTexp}
\eeq
This form, however, is less favorable for numerical reasons, thus we
have exclusively used~\eqref{eq:solgap_muTdirect} in the numerical
studies performed in section~\ref{ssec:solgap_calcmeth}.  Both forms are plotted in fig.~6.

\begin{figure}
\includegraphics[width=8cm]{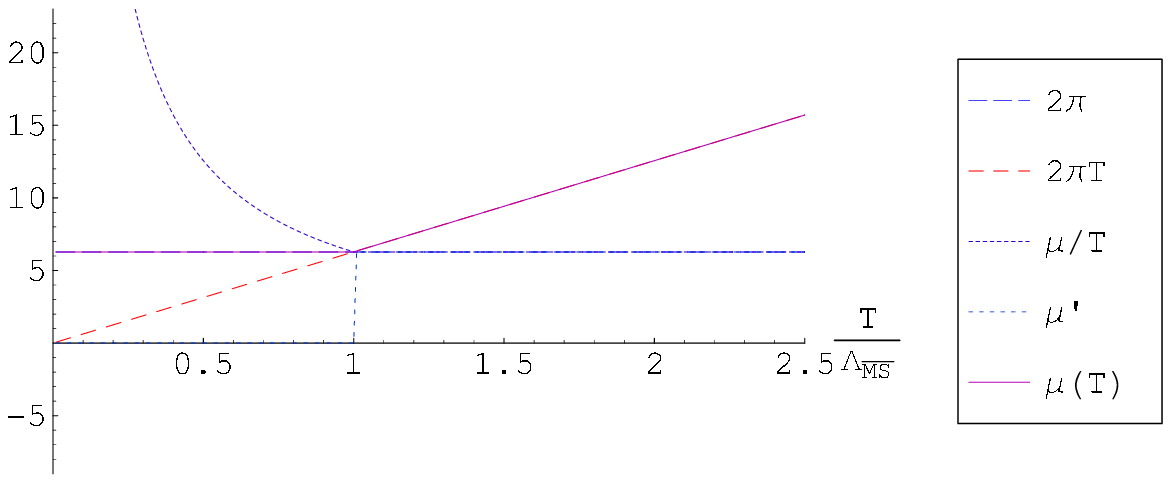}\\
\includegraphics[width=8cm]{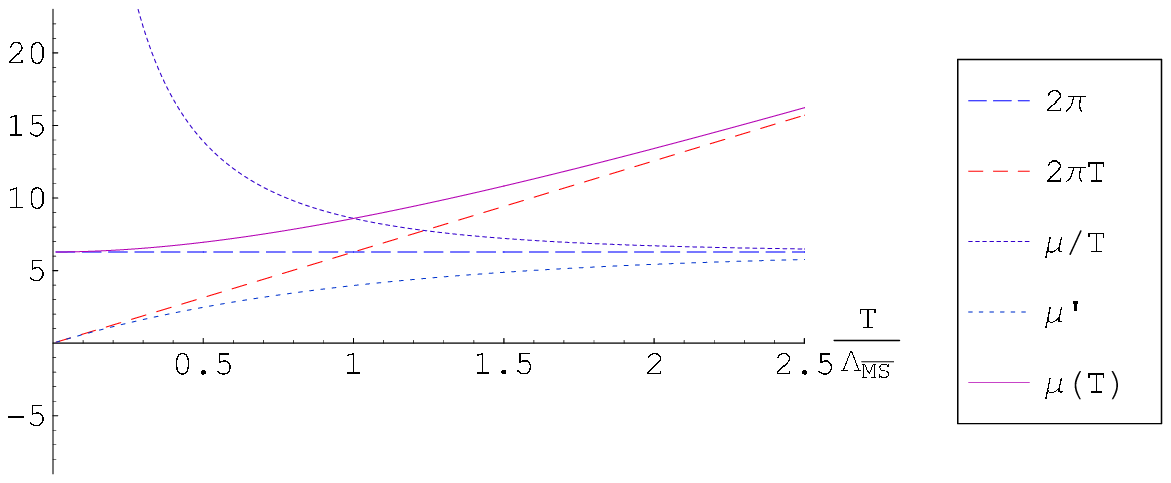}
\caption{The choices~\eqref{eq:solgap_muTdirect}
  and~\eqref{eq:solgap_muTexp} for the renormalization
  scale $\mu(T)$, displayed together with the asymptotics
  $\mu_{0,\mathrm{l}}=2\pi\Lambda_{\mathrm{QCD}}$,
  $\mu_{0,\mathrm{h}}=2\pi T$. In addition we show
  $\frac{\mu(T)}{T}$ and $\mu'=\frac{\d \mu}{\d T}$.}
\label{fig:solgap_muT}
\end{figure}

Another ansatz, used especially in functional calculations,~\cite{Maas:2005hs},
is a 't Hooft-like scaling
\beq
  g^2(\mu(T))\,T = \text{const}\,,
\eeq
which, at one-loop level, corresponds to an exponential growth
$\mu(T)=\mu_0\,\E^{\alpha T}$ with some positive constant $\alpha$.
This choice provides a smooth infinite-temperature limit, but does
not respect condition (II), and has not been used in the current article.

\subsection{Calculational Methods}
\label{ssec:solgap_calcmeth}

\noindent The gap equation~\eqref{eq:solgap_gapeq} is an implicit equation
for $m^*(T)$, which, in constrast to ``genuine'' integral equations, can be
solved independently for each temperature $T$. Our results have been
obtained in Mathematica by combining a numerical equation solver with
adaptive Gau\ss-Legendre integration.

The derivatives necessary to obtain anomaly and bulk viscosity (see
equations~\eqref{eq:solgap_Ar_fromw} and~\eqref{eq:solgap_bulkviscrednonpert})
can be done either numerically or analytically. The second way unfortunately
involves additional integrals, which can again only be evaluated
numerically. (See appendix~\ref{sec:solgap_app_anander} for details.)

While both methods are potentially susceptible to numerical problems,
they are of very different nature. Actually,  the results of both methods agree
remarkably well, inspiring confidence in the stability of the result.

All calculations directly involving $T$ have been performed on logarithmic
temperature scale. This allows direct implementation of logarithmic derivatives,
reduces numerical errors as compared to calculations on a linear scale and
enables one to reach significantly higher temperatures.

For all quantities under consideration, we could obtain asymptotic expressions
by expansion in the coupling $g^2$. In general, we use,~\cite{Laine:2005ai},
\begin{align}
  \frac1{g^2(\mu)} &\overset{2\text{-loop}}= 2b_0\,\ln\frac{\mu}{\Lambda_{\overline{\text{MS}}}}
  + \frac{b_0}{b_1}\,\ln\left( 2\,\ln\frac{\mu}{\Lambda_{\overline{\text{MS}}}} \right)\,,
  \label{eq:solgap_g2} \\
  \mu\,\frac{\d g^2}{\d \mu} &= \beta(g^2) \overset{2\text{-loop}}= \frac{\beta_0}{(4\pi)^2}\,g^4(\mu)
    + \frac{\beta_1}{(4\pi)^4}\,g^6(\mu)  \label{eq:solgap_beta}
\end{align}
with the coefficients
\begin{align}
  \beta_0 &\equiv -2\,(4\pi)^2\,b_0 = \frac{-22\,C_A+8T_f}{N} \nonumber \\
    &\overset{\mbox{\footnotesize pure SU(3)}}=\; -22\,, \\
  \beta_1 &\equiv -2\,(4\pi)^4\,b_1 = \frac{-68C_A^2+40C_AT_f+24C_fT_f}{N} \nonumber \\
    &\overset{\mbox{\footnotesize pure SU(3)}}=\; - 204
\end{align}
and the group-theoretical factors $C_A=N$ and $C_F=N^2-1$.
$T_f$ is equal to half the number of quark flavors and thus vanishes in pure gauge.
While the results in subsection~\ref{ssec:solgap_results} are given for the one-loop form
(easily obtained by setting $\beta_1=b_1=0$ in~\eqref{eq:solgap_g2}
and~\eqref{eq:solgap_beta}), there are only minor changes when switching
to the two-loop form.

\subsection{Results}
\label{ssec:solgap_results}

\noindent We now summarize the results obtained by numerically solving
the gap equation and the corresponding asymptotic expressions.

\subsubsection{Gribov Mass}

\begin{figure}
\raisebox{3cm}{(a)}
\includegraphics[width=7.5cm]{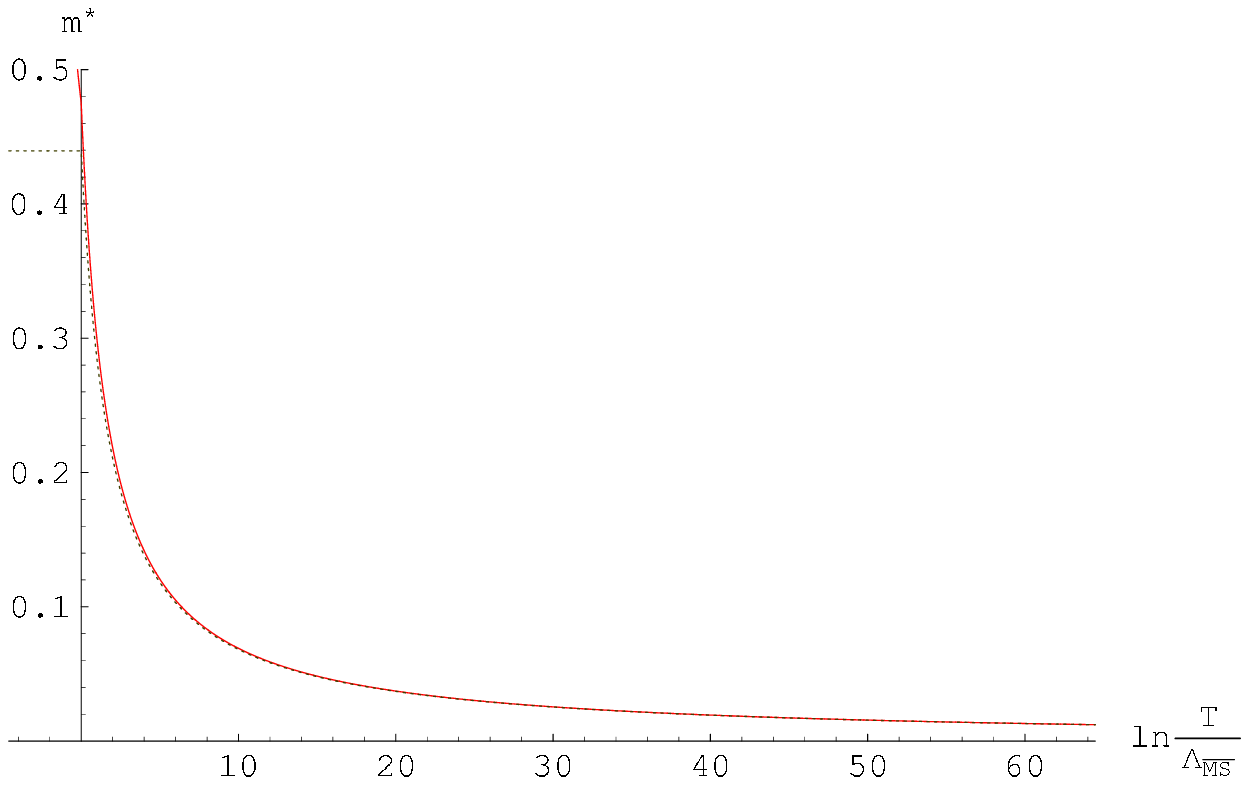} \\
\raisebox{3cm}{(b)}
\includegraphics[width=7.5cm]{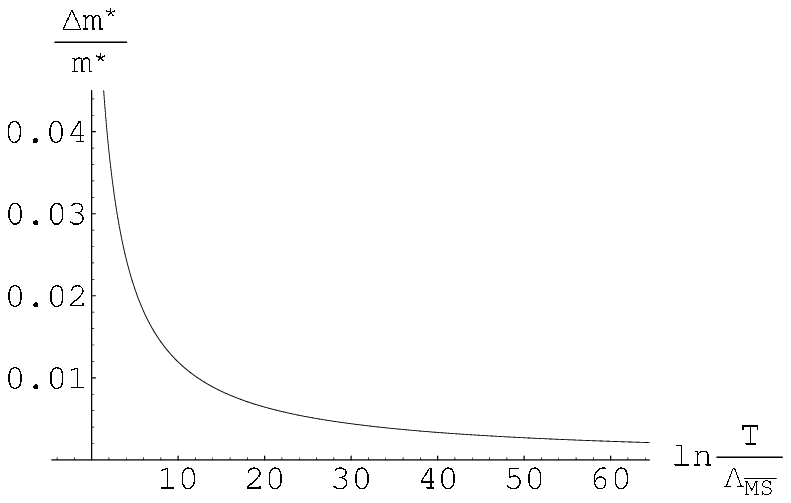}
\caption{The rescaled Gribov mass $m^*=\frac{m}T$: (a) {\color{red}solid} -- numerical solution,
 {\color{darkgray}dotted} -- asymptotic expression from~\eqref{eq:solgap_mstar_asy};
 (b) relative deviation
 $\Delta m^*_{\mathrm{rel}} = (m^*_{\mathrm{num}}-m^*_{\mathrm{asy}})/m^*_{\mathrm{asy}}$}
\label{fig:solgap_ms}
\end{figure}

\noindent Solving the gap equation yields the Gribov mass $m(T)$.
An expansion gives to leading order in $g^2$
\beq
  m^*(T) \sim { N   \over 2^{3/2}  \ 3 \  \pi } \ g^2(\mu)\,.
  \label{eq:solgap_mstar_asy}
\eeq
The numerical result and this asymptotic form are displayed in
fig.~\ref{fig:solgap_ms}. The agreement is excellent down to the phase
transition (below which the formalism is probably not applicable anyhow),
thus higher-order corrections to the Gribov mass are small.

\subsubsection{Free Energy and Pressure}

\begin{figure}
\raisebox{3cm}{(a)}
\includegraphics[width=7.5cm]{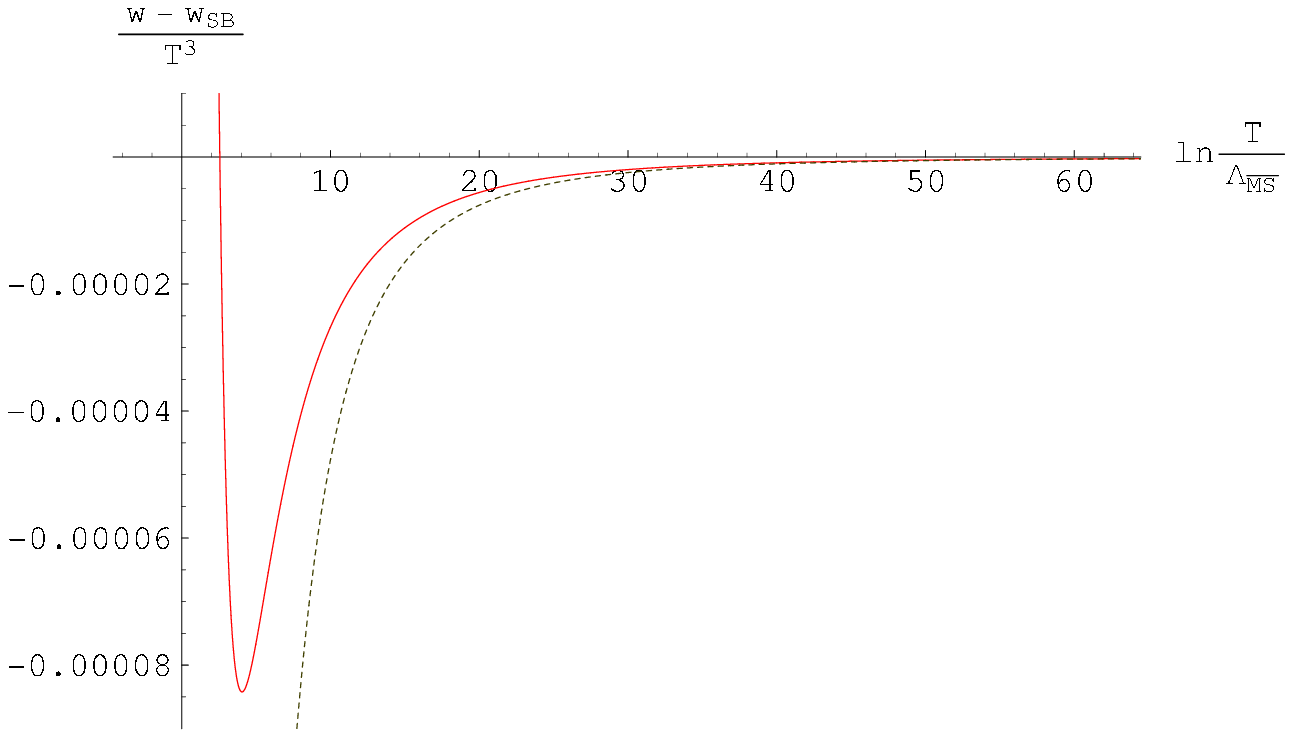} \\
\raisebox{3cm}{(b)}
\includegraphics[width=7.5cm]{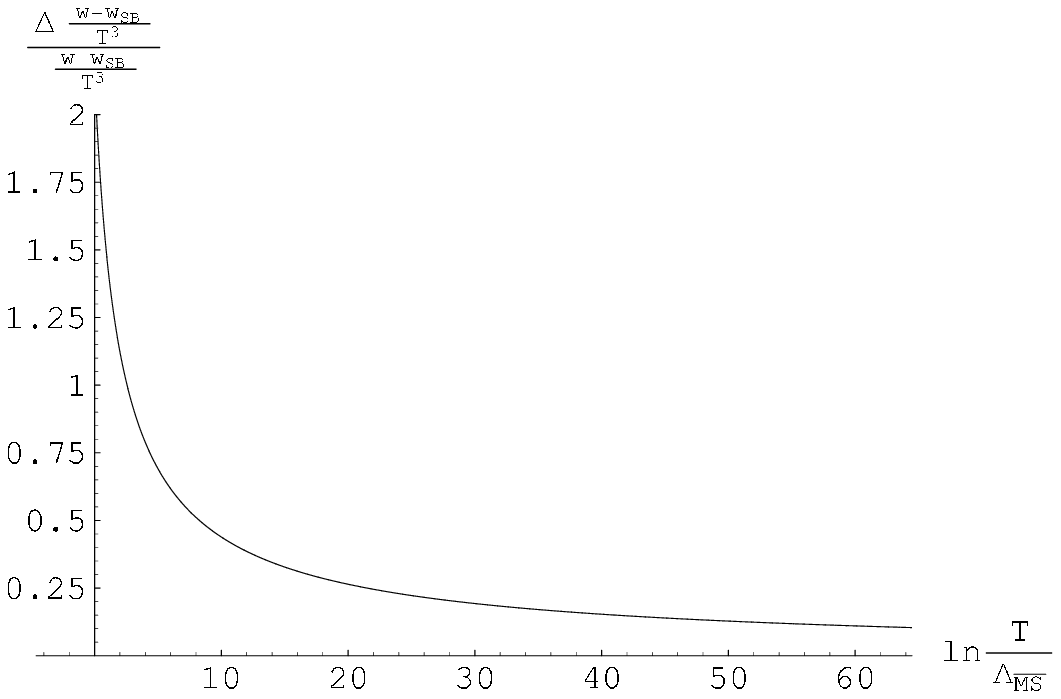}
\caption{The rescaled reduced free energy $w_r-w_{\text{r,SB}}$: (a) {\color{red}solid} -- numerical solution,
 {\color{darkgray}dotted} -- asymptotic expression from~\eqref{eq:solgap_w_asy};
 (b) relative deviation
 $\Delta w_{\mathrm{r,rel}} = (w_{\mathrm{r,num}}-w_{\mathrm{r,asy}})/
   (w_{\mathrm{r,asy}}-w_{\text{r,SB}})$}
\label{fig:solgap_w}
\end{figure}

\noindent For the pressure $p$ and the free energy $w$ we obtain
\begin{align}
  \frac{p}{T^4}  \equiv \frac{w}{T^3}
    &= (N^2-1)\,\left[ \frac{3}{2N}\,\frac{{m^*}^4}{g^2(\mu)} + \frac1{3\pi^2 T^4}\,K(m) \right]\,, \\
  K(m) &:= \int_0^{\infty} \frac{\d k}{E(k)}\,\frac{k^4-m^4}{\E^{\beta E(k)}-1}\,,
\end{align}
with $k=\norm{\V{k}}$. An expansion for $K(m)$ is not completely straightforward
due to a nonanalyticity in $m^4$, but, as shown in~\cite{Zwanziger:2006sc}, it can
be performed and yields the asymptotic expression
\beq
  w \sim (N^2 -1) {\pi^2 \over 45 } \ \left( 1 - 
 {5 \over 18} \left( {N g^2 \over 4 \pi^2 }\right)^3 
    \right) \ T^3\,.
  \label{eq:solgap_w_asy}
\eeq
The full solution and the asymptotic form are given in fig.~\ref{fig:solgap_w}, where
we have subtracted the Stefan-Boltzmann part, denoted by $w_{\text{SB}}$. In contrast
to the case of $m$, higher-order corrections are obviously not small for $w$ since
agreement between the full (numerical) and the asymptotic result is not good below
$T\approx 10^6\,\Lambda_{\bar{MS}}$.

\medskip

It is instructive to see that $K(m)$ can also be evaluated by using an intermediate
cutoff.  While more cumbersome, this method allows us to identify contributions
from different scales and thus gives some idea how to relate this result to the one
obtained by effective theory approaches (see section~\ref{ssec:solgap_effField}).

To do this, we introduce a cutoff $\Lambda$ with $m\ll \Lambda \ll T$, which
separates contributions from the scale $m\sim g^2T$ and from the scale $2\pi T$.
Doing so, we obtain
\beq
  K(m) = \underbrace{\int_0^{\Lambda} \frac{\d k}{E(k)}\,\frac{k^4-m^4}{\E^{\beta E(k)}-1}}_{K_1}
  + \underbrace{\int_{\Lambda}^{\infty} \frac{\d k}{E(k)}\,\frac{k^4-m^4}{\E^{\beta E(k)}-1}}_{K_2},
\eeq
where
\begin{align}
  K_1 &= m^4\int_0^{\Lambda/m} \d x\,\frac{x^4-1}{u(x)\,(\E^{\beta m u(x)}-1)} \nonumber \\
    &\approx m^3T\int_0^{\Lambda/m} \d x\,\frac{x^4-1}{u^2(x)} 
  = m^3T\int_0^{\Lambda/m} \d x\,x^2\frac{x^4-1}{x^4+1} \nonumber \\
  &=m^3T\left\{ \int_0^{\Lambda/m} \d x\,x^2 - 2 \int_0^{\Lambda/m} \frac{x^2}{x^4+1} \,\d x\right\}\,.
\end{align}
The first integral is trivial; for the second we can replace the upper limit $\Lambda/m$ by
$\infty$ and apply residue calculus to obtain
\beq
  K_1 = \frac{T}3\,\Lambda^3 - \frac{\pi}{\sqrt2}\,m^3\,T\,.
\eeq

For $K_2$ we obtain
\begin{align}
  K_2 & \approx \int_{\Lambda}^{\infty} \frac{\d k}{k}\,\frac{k^4}{\E^{\beta k}-1}
  = T^4 \int_{\Lambda/T}^{\infty} \d y\, \frac{y^3}{\E^y-1} \nonumber \\
  &= T^4 \left\{ \int_0^{\infty} \d y\, \frac{y^3}{\E^y-1}  - \int_0^{\Lambda/T} \d y\, \frac{y^3}{\E^y-1} \right\}\,.
\end{align}
The first integral is the well-known Planck integral.  In the second one we can again
expand the exponential, since $y\le\frac{\Lambda}{T}\ll 1$, and obtain
\beq
  K_2 \approx T^4\,\left\{ \frac{\pi^4}{15} - \int_0^{\Lambda/T} \d y\, y^2 \right\}
  = T^4\,\frac{\pi^4}{15} - \frac{T}3\,\Lambda^3\,.
\eeq
This gives
\beq
  K = K_1+K_2 = T^4\,\frac{\pi^4}{15} - \frac{\pi}{\sqrt2}\,m^3\,T\,.
\eeq
The cutoff-dependent parts in $K_1$ and $K_2$ precisely cancel, leaving
a clear separation of the Stefan-Boltzmann contribution from $k\sim T$
and the contribution from the scale $k\sim m\sim g^2T$.

\subsubsection{Energy and Anomaly}

\noindent From the free energy or the pressure, we can calculate the rescaled anomaly via
\beq
  A_{\mathrm{r}} = \frac{e-3p}{T^4} = T\,\frac{\mathrm{d}}{\mathrm{d}T} \frac{p}{T^4}
  = \frac{\mathrm{d}}{\mathrm{d}(\ln\frac{T}{\Lambda})} \frac{p}{T^4}\,,
  \label{eq:solgap_Ar_fromw}
\eeq
(with some arbitrary scale $\Lambda$) since from~\eqref{eq:solgap_thermoform}, we have
\begin{align}
  T\,\frac{\mathrm{d}}{\mathrm{d}T} \frac{p}{T^4}
  &\equiv T\,\frac{\mathrm{d}}{\mathrm{d}T} \frac{w}{T^3}
  = \frac{\frac{\partial w}{\partial T}}{T^2} - 3\,\frac{w}{T^3} \nonumber \\
  &=\frac{T^2\frac{\partial w}{\partial T}-3wT}{T^4}
  =\frac{e-3p}{T^4}\,.  
\end{align}
It is also obvious that the energy can be directly obtained from the anomaly
by using the relation $e=3p+A$. Thus we do not show separate graphs for $e$.
From~\eqref{eq:solgap_Ar_fromw} it is also clear that \emph{all} deviations
from the Stefan-Boltzmann pressure $p_{\text{r,SB}}=\frac{\pi^4}{15} $
are encoded in the anomaly, since integration gives
\beq
  p_{\text{r}}(T) = p_{\text{r,SB}} - \int_T^{\infty} \frac{A_{\text{r}}(T')}{T'}\,\d T'\,.
\eeq

\begin{figure}
\raisebox{3cm}{(a)}
\includegraphics[width=7.5cm]{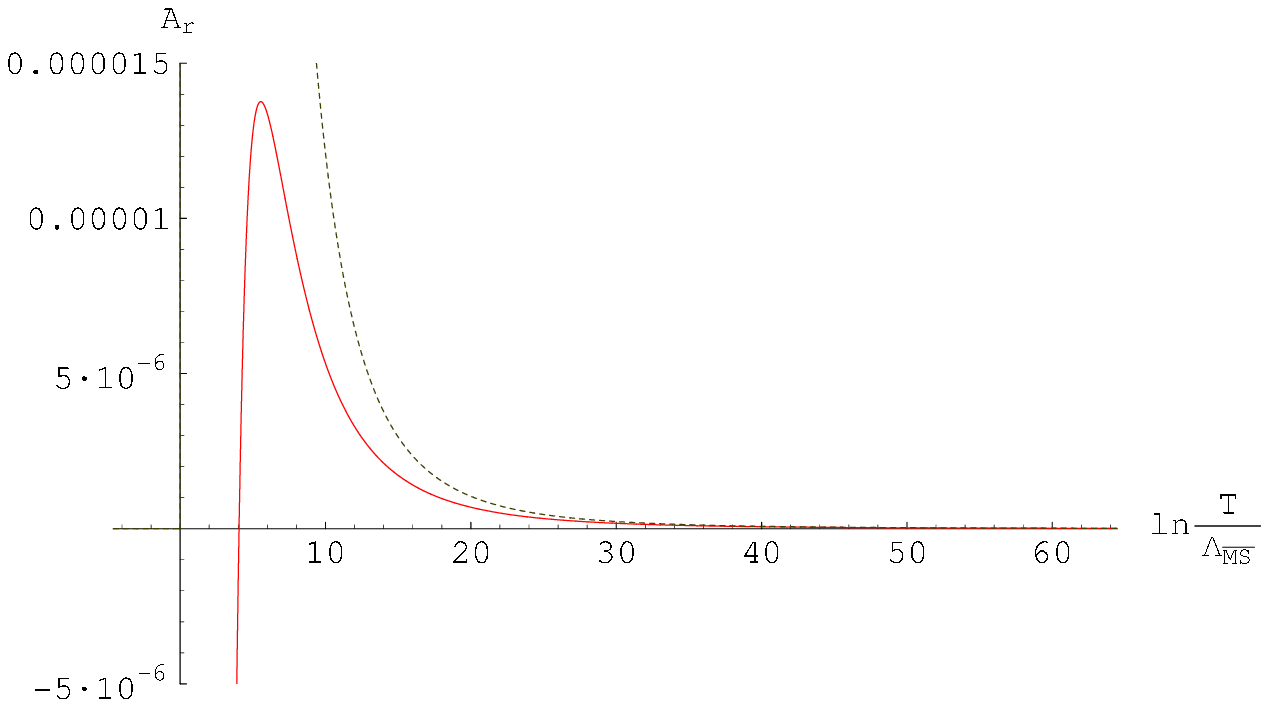} \\
\raisebox{3cm}{(b)}
\includegraphics[width=7.5cm]{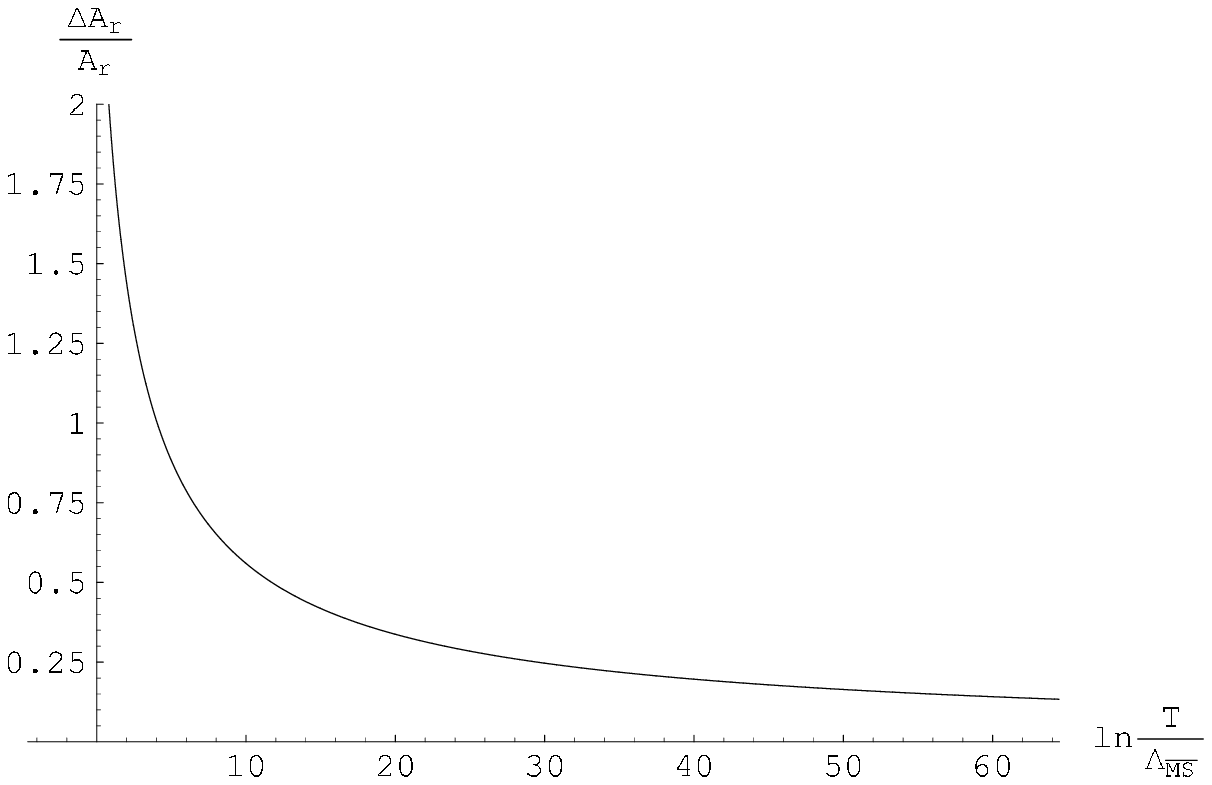}
\caption{The rescaled anomaly $A_r$: (a) {\color{red}solid} -- numerical solution,
 {\color{darkgray}dotted} -- asymptotic expression from~\eqref{eq:solgap_Ar_asy};
 (b) relative deviation
 $\Delta A_{\mathrm{r,rel}} = (A_{\mathrm{r,num}}-A_{\mathrm{r,asy}})/A_{\mathrm{r,asy}}$}
\label{fig:solgap_Ar}
\end{figure}

From~\eqref{eq:solgap_w_asy},~\eqref{eq:solgap_Ar_fromw}, and \eqref{eq:solgap_beta} we obtain
\beq
  A_{\mathrm{r}} \sim -(N^2 - 1)\frac{N^3}{3456\,\pi^4}\,g^4(\mu)
        \beta(g^2) \frac{T}{\mu}\,\frac{\d \mu}{\d T}\,.
    \label{eq:solgap_Ar_asy}
\eeq
for the asymptotic expansion. The numerical result and the
asymptotic form are shown in fig.~\ref{fig:solgap_Ar}. Again
higher-order corrections are large except for extremely high
temperatures.

\subsubsection{Bulk Viscosity}

\noindent In formula~\eqref{eq:solgap_viscos_kharzeev} there is one ambiguity,
the choice of the scale $\omega_0$. According to~\cite{Karsch:2007jc} a reasonable
range of values is $\omega_0=(0.5\div1.5)$~GeV. Neglecting the perturbative
contribution from $\eps_V$, we obtain
\beq
  \zeta = \frac1{9\,\omega_0}\,T^5\,\frac{\d}{\d T} \frac{e-3p}{T^4}\,.
  \label{eq:solgap_bulkviscnonpert}
\eeq
The rescaled bulk viscosity is given by
\beq
  \zeta_{\text{r}} = \frac1{9\,\omega_0}\,T\,\frac{\d}{\d T} A_{\text{r}}\,.
  \label{eq:solgap_bulkviscrednonpert}
\eeq

\begin{figure}
\raisebox{3cm}{(a)}
\includegraphics[width=7.5cm]{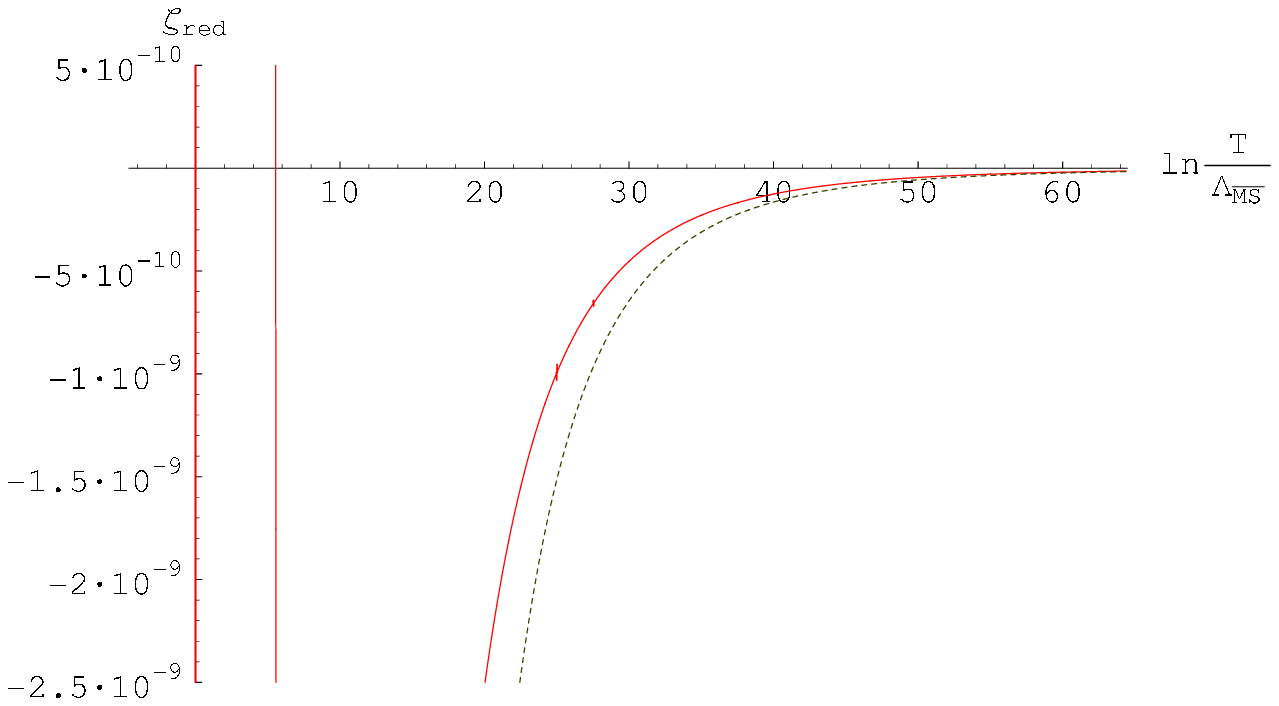}\\
\raisebox{3cm}{(b)}
\includegraphics[width=7.5cm]{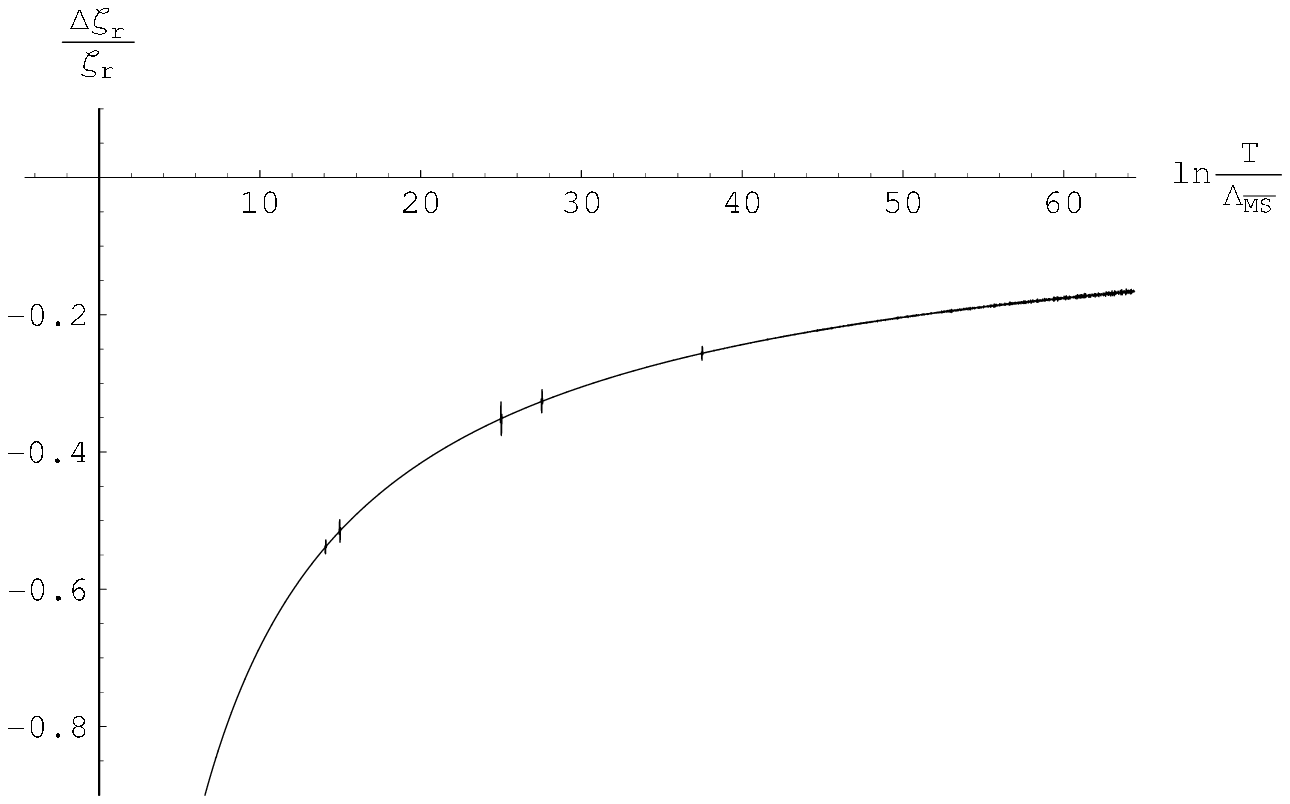}
\caption{The rescaled bulk viscosity: (a) {\color{red}solid} -- numerical
 solution $\zeta_{\mathrm{r,num}}$, {\color{darkgray}dotted} -- asymptotic expression
 $\zeta_{\mathrm{r,asy}}$ from~\eqref{eq:solgap_Ar_asy}; (b) relative deviation
  $\Delta \zeta_{\mathrm{r,rel}}
  =(\zeta_{\mathrm{r,num}}-\zeta_{\mathrm{r,asy}})/\zeta_{\mathrm{r,asy}}$}
\label{fig:solgap_viscos}
\end{figure}

In the asymptotic expression, correction terms originating from
$\frac{T}{\mu}\,\frac{\d \mu}{\d T}$ become quite complicated.
Since they are relatively unimportant for reasonable choice of $\mu(T)$
(and even vanish identically for the simple form~\eqref{eq:solgap_muTdirect})
we only give the simplified expression, where we have set
$\frac{T}{\mu}\,\frac{\d \mu}{\d T}=1$,
\begin{align}
  \zeta_{\text{r}}
  &\sim - \frac1{9\,\omega_0}\,\frac{N^3\,(N^2-1)}{3456\,\pi^4} \nonumber \\
  &\qquad \left\{ 2\,g^2(\mu)\,\,\beta(g^2) + g^4(\mu)\,
     \frac{\d\beta(g^2)}{\d g^2}\right\}\,\beta(g^2).
\end{align}
The full expression is derived in appendix~\ref{sec:solgap_app_viscos}.
Graphs for the numerical solution and the asymptotic expression are shown in
fig.~\ref{fig:solgap_viscos} for the choice $\omega_0=5\Lambda_{\overline{\text{MS}}}$.

The behaviour close to $T=\Lambda_{\bar{MS}}$ is strongly influenced
by the choice of $\mu(T)$. Apart from that however the viscosity $\zeta_{\text{r}}$
rises significantly when the temperature approaches the critical temperature
from above, in agreement with~\cite{Karsch:2007jc}.

\section{Discussion}
\label{sec:solgap_discussion}

\subsection{Access to the nonperturbative sector}
\label{ssec:solgap_access_nonpert}

\noindent Various results make clear that finite-temperature QCD contains in principle 
a perturbatively accessible sector, which, starting at order $(g^2T)^3$, interacts
with a genuine nonperturbative sector. At least formally an expansion in powers
of the coupling $g$ is possible also for non-perturbative contributions.

According to Gribov's confinement scenario,~\cite{Gribov:1977wm, Zwanziger:1991gz, Zwanziger:1992}, the vicinity of the
Gribov horizon dominates the nonperturbative aspects of the theory. So correctly
taking into account this region should give access to the nonperturbative sector of the
theory at high temperatures also.   
Indeed, the cutoff at the Gribov horizon employed in this article gives a finite nonperturbative
contribution to the free energy at order $g^6$, where the nonperturbative sector of the
theory begins to spoil direct perturbative approaches.

The nonperturbative sector (described by MQCD in the picture of section~\ref{ssec:solgap_effField})
is also accessible to lattice calculations. Comparison of our analytic result~\eqref{eq:solgap_w_asy}
with the lattice expressions~\eqref{eq:solgap_lattice_npcoeff1} and~\eqref{eq:solgap_lattice_npcoeff2}
gives
\begin{align}
  w_{\text{np, analyt}}^{(6)} &= - \frac{(N^2\!-\!1)\,N^3}{10\,368\,\pi^4} g^6 T^3\,, 
    \label{eq:solgap_resultanalyt} \\
  w_{\text{np, lattice}}^{(6)} &= - \frac{(N^2\!-\!1)\,N^3}{1\,280\,\pi^4}(1 \pm 4) g^6 T^3\,.
    \label{eq:solgap_resultlattice}
\end{align}
These results are compatible, though the errors of the lattice calculations are too large at the moment to allow a definite statement about the quality of agreement.

\subsection{Convergence of the series?}
\label{ssec:solgap_convseries}

\noindent As already mentioned in section~\ref{ssec:solgap_compare}, the convergence
of perturbation series is extremely poor for temperatures $\mathcal{O}(\text{GeV})$ or below.
As discussed in~\cite{Blaizot:2003iq}, this can be traced back to the poor convergence of
contributions from the EQCD sector, which begin to contribute at order $g^3$.

\begin{figure}
\includegraphics[width=8cm]{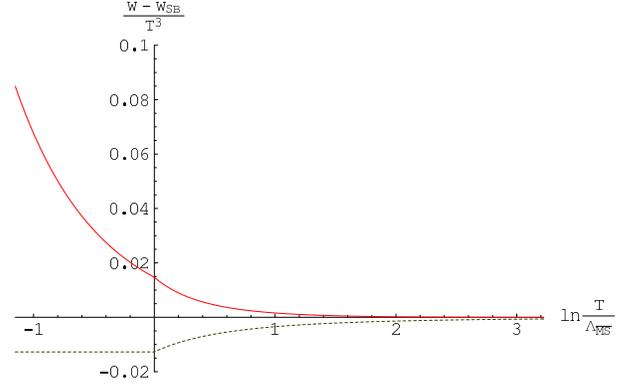}
\caption{The rescaled free energy in the low-temperature
 region ({\color{red}solid} -- numerical solution,
 {\color{darkgray}dotted} -- asymptotic expression
 from~\eqref{eq:solgap_Ar_asy})}
\label{fig:solgap_wr_lowT}
\end{figure}

\begin{figure}
\includegraphics[width=8cm]{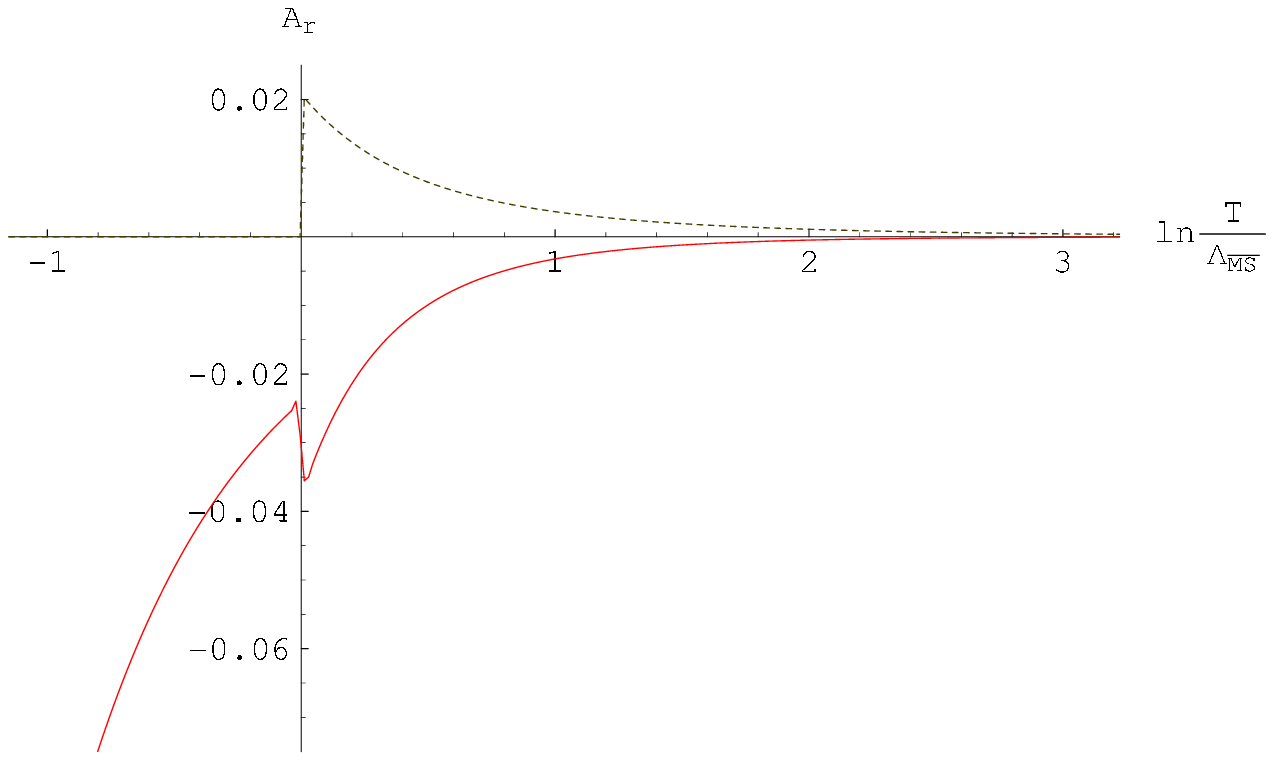}
\caption{The rescaled anomly $\frac{e-3p}{T^4}$ in the
 low-temperature region ({\color{red}solid} -- numerical
 solution, {\color{darkgray}dotted} -- asymptotic expression
 from~\eqref{eq:solgap_Ar_asy})}
\label{fig:solgap_Ar_lowT}
\end{figure}

\begin{figure}
\includegraphics[width=8cm]{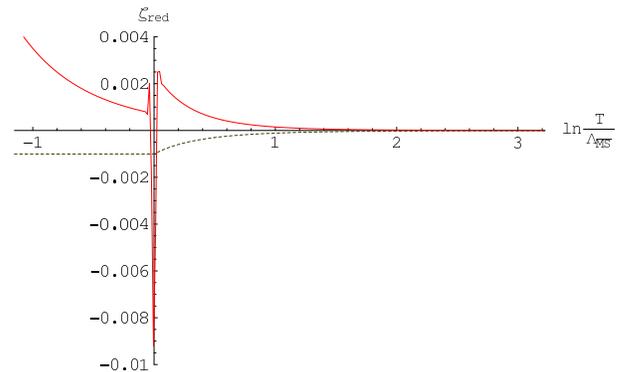}
\caption{The rescaled bulk viscosity in the
 low-temperature region ({\color{red}solid} -- numerical
 solution, {\color{darkgray}dotted} -- asymptotic expression
 from~\eqref{eq:solgap_Ar_asy})}
\label{fig:solgap_viscr_lowT}
\end{figure}

A similar behaviour seems to be true for the contribution from MQCD. While a formal expansion
in $g$ is possible (and for very high temperatures $T\ge 10^{10}\text{GeV}$ the agreement
is reasonably good), the expansion has little to do with the full result for low temperatures.
From the low-temperature graphs displayed in figs.~\ref{fig:solgap_wr_lowT}
to~\ref{fig:solgap_viscr_lowT} it is likely that higher-order corrections cannot be
small compared to the leading term.

This suggests that either the convergence is extremely poor or that
there is even no convergence at all. The latter would not be completely
unexpected. It is well known that in quantum field theory series obtained
by expansion in the coupling are rarely convergent, but at best asymptotic
(and usually even this cannot be proven).

Assuming that expansion in $g$ of the QCD free energy yields a divergent
asymptotic series would give the following scenario: For each temperature
$T$ one can expect that an ``optimal order'' $n$ exists beyond which the
series leaves the ``path of apparent convergence''. For low temperatures
and thus large couplings this order may be so small that no partial sum
of the perturbation series can serve as a satisfactory approximation.

\subsection{Further Steps}
\label{ssec:solgap_furthersteps}

\noindent Our studies leave open several questions that are worth further
investigation. While the analytic result~(\ref{eq:solgap_resultanalyt}) is
compatible with the lattice result~(\ref{eq:solgap_resultlattice}), a reduction of
the errors of the latter would be highly desirable. (Unfortunately the errors would
have to be reduced about one order of magnitude, a task which would
require considerable computational resources.)

On the analytic side, higher-order calculations in the semi-perturbative formalism
could help to further clearify the connection of our approach to the sequence
of theories discussed in subsection~\ref{ssec:solgap_effField}. They could
also help to reveal if~(\ref{eq:solgap_resultanalyt}) is indeed the full contribution
to order $g^6$ from the magnetostatic sector or if there are additional
contributions from (formally) higher orders as well, which are not present
in the lowest-order approximation to the gap equation~\eqref{eq:solgap_gapeq}.

Such calculations could also shed some more light on the question of (apparent)
convergence, as just discussed in subsection~\ref{ssec:solgap_convseries}.
Of course also determination of the full $g^6$ contribution to the free energy
(conceptually possible in the framework of effective theories) would be very helpful
for further statements about convergence issues.

Our results indicate that the semi-perturbative method of calculation  is reliable only
at extremely high $T$.  A more advanced approach to the nonperturbative sector could involve
solving the Dyson-Schwinger equation for the system with local action and auxiliary
fermi and bose ghost pairs~\cite{Zwanziger:2006sc}
or, alternatively, studying bound-state equations for glueballs in MQCD.
Those objects, which determine higher-order contributions from this sector are
closely related (though not strictly identical) to chromomagnetic glueballs
in the four-dimensional theory.

\section{Summary and Outlook}
\label{klsec:solgap_summary}

\noindent We first presented a short synopsis on current problems
in thermal QCD, where we also reviewed several methods to
approach them, including consecutive effective field theories, as
proposed in~\cite{Braaten:1995cm}.

Then, using the local action proposed in~\cite{Zwanziger:2006sc},
we obtained nonperturbative contributions to several thermodynamic
observables, including free energy, anomaly and bulk viscosity. Being
directly related to other nonperturbative approaches (like magnetostatic
QCD), this method provides a framework for purely analytical studies
beyond the limits of thermal perturbation theory, without the necessity
of lattice calculations (neither in the full nor in an effective theory.)

While for each quantity we were able to obtain the leading coefficient
of the expansion in the coupling $g$, we also noticed that higher-order
corrections have to be large in order to accomodate the numerical results.
This is one more sign that, in thermal QCD, expansion in the coupling
cannot be expected to give reliable approximations except for extremely
high temperatures.

While results are promising, further investigations are necessary to
clearify the limits of this method, illuminate further the relation to other
approaches and perhaps extract general information about the quality
of series expansions in thermal QCD.

\bigskip

{\bf Acknowledgements}\\

\noindent K.L was supported by the Doktoratskolleg \emph{Hadronen
im Vakuum, in Kernen und Sternen} (FWF DK W1203-N08)  and by the
\emph{Graz Advanced School of Science} (NAWI-GASS). He would like
to  express his thanks for the hospitality of New York University (NYU),
where most of this work was done, and to Reinhard Alkofer for his support,
in particular for making possible the stay at NYU.

The authors are especially grateful to Reinhard Alkofer for valuable
comments and to Roman Scoccimarro for developing an early
version of the numerical program used to solve the gap equation.
They would like to thank Fritjof Karsch and J\"urgen Engels for providing
lattice data from~\cite{Boyd:1996bx} and Mikko Laine for making them aware
of references~\cite{Hietanen:2004ew, Hietanen:2006rc, DiRenzo:2006nh,
Schroder:2002re, Schroder:2003uw}.

\appendix

\section{The Anomaly from an Analytic Derivative}
\label{sec:solgap_app_anander}

\noindent We now show how the anomaly can be obtained without
the necessity for numerical differentiation. In the following,
we always understand $g^2=g^2(\mu)$ and $\mu=\mu(T)$.

One-loop expansion of the gap equation gives for the rescaled pressure
\beq
  \frac{p}{T^4} \equiv \frac{w}{T^3} = (N^2-1)\,\left[ \frac{3}{2N}\,\frac{{m^*}^4}{g^2(\mu)} + \frac1{3\pi^2}\,K(m^*) \right]
\eeq
with $K(m^*) = {m^*}^4\,I(m^*)$ and
\beq
  I(m^*) := \int_0^{\infty} \frac{\d x}{u(x)}\,\frac{x^4-1}{\E^{m^*u(x)}-1}\,,
\eeq
using the reduced dispersion relation~\eqref{eq:solgap_III_reddisprel}.
From this, we obtain
\begin{align}
  A_{\mathrm{r}} & =  T\,\,\frac{\d}{\d T}\frac{p}{T^4} \nonumber \\
  &= (N^2-1) \left[ \frac{3}{2N}\,T\,\frac{\d}{\d T} \frac{{m^*}^4}{g^2}  
  + \frac1{3\pi^2}\,T\,\frac{\partial K(m^*)}{\partial T} \right] \nonumber \\
&= 3\,\frac{N^2-1}{2N} \left[ -\frac{{m^*}^4}{g^4}\,T\,\frac{\d g^2}{\d T}
+ \frac{4\,{m^*}^3}{g^2}\,T\,\frac{\d m^*}{\d T}  \right] \nonumber \\
&\qquad + \frac{N^2-1}{3\pi^2}\,T\,\frac{\partial K(m^*)}{\partial T}\,.
  \label{eq:solgap_app_anomaly}
\end{align}
The derivative of $g^2$ with respect to $T$ be easily calculated using the $\beta$-function,
\beq
  T\,\frac{\d(g^2)}{\d\mu} = \frac{T}{\mu}\,\mu\,\frac{\d(g^2)}{\d\mu} = \frac{T}{\mu}\,\beta(g^2)\,.
  \label{eq:solgap_betag2}
\eeq
From this, we obtain
\begin{align}
  T\,\frac{\d g^2(\mu)}{\d T} &= \frac{T}{\mu}\,\mu \, \frac{\d g^2(\mu)}{\d \mu}\,\frac{\d\mu(T)}{\d T}
  = \beta(g^2(\mu))\, \frac{T}{\mu}\,\frac{\d\mu(T)}{\d T}\,.
\end{align}
For the last term in~\eqref{eq:solgap_app_anomaly} we have
\begin{align}
  T\,\frac{\partial K(m^*)}{\partial T} &= T\,\frac{\partial m^*}{\partial T}\,\frac{\partial K(m^*)}{\partial m^*}\, \\
  \frac{\partial K(m^*)}{\partial m^*} &= 4\,{m^*}^3\,I(m^*) - {m^*}^4\,L(m^*)\, \\
  L(m^*) &= \int_0^{\infty} \d x\, \frac{(x^4-1)\,\E^{m^*u(x)}}{(\E^{m^*u(x)}-1)^2}\,.
\end{align}
We obtain $T\,\frac{\partial m^*}{\partial T}$ by differentiating the gap equation
$f \left( m^* \right) = y(T)$ with respect to $T$, where
\begin{align}
    f \left( m^* \right)  &= \frac12\ln\frac1{m^*} + \int_0^{\infty} \frac{\d x}{u(x)}\,\frac1{\E^{m^*u(x)}-1}\,, \\
   y(T) &= \frac{3\pi^2}{N\,g^2(\mu)} - \frac14\,\ln\frac{\E\mu^2(T)}{2\,T^2}\,.
\end{align}
Since the left-hand side depends on $T$ only implicitly via $m^*$,
we have
\begin{align}
    \frac{\d}{\d T} f(m^*) &= \frac{\d f}{\d m^*}\,\frac{\d m*}{\d T}
    = \left[ - \frac12\,\frac1{m^*}  - J(m^*) \right]\,\frac{\d m^*}{\d T} \,, \\
  J(m^*) &= \int_0^{\infty} \d x\,\frac{\E^{m^*u(x)}}{\left(\E^{m^*u(x)}-1\right)^2}\,.
\end{align}
Differention of the right-hand side yields
\begin{align}
  \frac{\d y(T)}{\d T} &= \frac{\d}{\d T} \left[
    \frac{3\pi^2}{N\,g^2(\mu)} - \frac14\,\ln\frac{\E\mu^2(T)}{2\,T^2} \right] \nonumber \\
    &= - \frac{3\pi^2}{N\,g^4(\mu(T))}\,\frac{\d g^2(\mu)}{\d T}
    - \frac{\d}{\d T} \frac12\,\ln\left( \sqrt{\frac{\E}{2}} \frac{\mu(T)}{T} \right) \nonumber \\
    &= - \frac{3\pi^2}{N\,g^4(\mu(T))}\,\frac1{\mu(T)}\,\mu(T)\frac{\d g^2(\mu)}{\d T} \nonumber \\
    &\quad-\frac12\,\frac{T}{\mu(T)} \left( \frac{\frac{\d\mu}{\d T}}{T} - \frac{\mu}{T^2}\right) \nonumber \\
    & =  - \frac1{\mu(T)}\,\left\{ \frac{3\pi^2}{N}\,\frac{\beta(g^2)}{g^4(\mu)}\,
  \frac{\d\mu}{\d T} + \frac12\left( \frac{\d \mu}{\d T} - \frac{\mu}{T} \right) \right\}\,.
\end{align}
The second term inside the curly brackets is a measure for the deviation from the
asymptotic behaviour; in the asymptotic regime with $\mu(T)=2\pi T$, the above
expression simplifies to
\beq
  \left. \frac{\d y(T)}{\d T} \right|_{\mathrm{asympt}} =
  - \frac{3\pi^2}{N}\,\frac1{T}\,\frac{\beta(g^2(2\pi T))}{g^4(2\pi T)} \,.
\eeq
Collecting our results, we obtain
\begin{align}
  T\,\frac{\d m^*}{\d T} &= \frac{T\,\frac{\d y}{\d T}}{-\frac1{2m^*} - J(m^*)} \nonumber \\
  &= \frac{T}{\mu(T)}\,\frac1{\frac1{2m^*}+J(m^*)} \nonumber \\
  &\cdot \left\{ \frac{3\pi^2}{N}\,\frac{\beta(g^2)}{g^4}\,
  \frac{\d\mu}{\d T} + \frac12\left( \frac{\partial\mu}{\partial T} - \frac{\mu}{T} \right) \right\}\,,
\end{align}
and the rescaled anomaly is given by
\begin{align}
 A_{\mathrm{r}}
  &= 3\,\frac{N^2-1}{2N} \left[ -\frac{{m^*}^4}{g^4}\,\beta(g^2)\, \frac{T}{\mu}\,\frac{\d\mu}{\d T}
  + \frac{4\,{m^*}^3}{g^2}\,T\,\frac{\d m^*}{\d T}  \right] \nonumber \\
  &\quad + \frac{N^2-1}{3\pi^2}\,\left( 4\,{m^*}^3\,I(m^*) - {m^*}^4\,L(m^*) \right) \, T\,\frac{\d m^*}{\d T}\,.
\end{align}

\bigskip

\section{Asymptotic Bulk Viscosity}
\label{sec:solgap_app_viscos}

\noindent Keeping the full dependence on $\mu(T)$, we obtain for the bulk
viscosity~\eqref{eq:solgap_bulkviscnonpert}, again making use
of~\eqref{eq:solgap_betag2},
\begin{align}
  \zeta &= \frac1{9\,\omega_0}\,T^5\,\frac{\d}{\d T} \frac{e-3p}{T^4}  \nonumber \\
  &\sim -\frac1{9\,\omega_0}\,\frac{N^3\,(N^2-1)}{3456\,\pi^4}\,
     T^5\,\frac{\d}{\d T} g^4(\mu)\,\beta(g^2) \frac{T}{\mu}\,\frac{\d\mu}{\d T}  \nonumber \\
  &\sim - \frac1{9\,\omega_0}\,\frac{N^3\,(N^2-1)}{3456\,\pi^4}\,T^5  \nonumber \\
  &\quad  \Bigg\{\left[ 2\,g^2(\mu)\,\,\beta(g^2) + g^4(\mu)\,\frac{\d\beta(g^2)}{\d g^2}\right]
     \frac{\d (g^2)}{\d\mu}\,\frac{\d\mu}{\d T}\,\frac{T}{\mu}\,\frac{\d\mu}{\d T}  \nonumber \\
  &\qquad+ g^4(\mu)\,\beta(g^2)\,\left[ \frac1{\mu}\,\frac{\d\mu}{\d T}
    - \frac{T}{\mu^2}\,\left( \frac{\d\mu}{\d T} \right)^2 + \frac{T}{\mu}\,\frac{d^2\mu}{\d T^2}\right] \Bigg\}  \nonumber \\
 &\sim  - \frac1{9\,\omega_0}\,\frac{N^3\,(N^2-1)}{3456\,\pi^4}\,T^4\,g^2(\mu)\,\beta(g^2)  \nonumber \\
  &\quad  \Bigg\{\left[ 2\beta(g^2) + g^2(\mu)\,\frac{\d\beta(g^2)}{\d g^2}\right]
     \left(\frac{T}{\mu}\,\frac{\d\mu}{\d T} \right)^2  \nonumber \\
  &\qquad+ g^2(\mu)\,\left[ \frac{T}{\mu}\,\frac{\d\mu}{\d T} -\left(\frac{T}{\mu}\,\frac{\d\mu}{\d T}\right)^2
     + \frac{T}{\mu^2}\,\frac{d^2\mu}{\d T^2}\right] \Bigg\}\,.
 \end{align}


\end{document}